\documentclass[fleqn,usenatbib]{mnras}

\usepackage{booktabs,caption}
\usepackage{arydshln}
\usepackage[flushleft]{threeparttable}
\usepackage{graphicx} 
\usepackage{subcaption} 

\usepackage[T1]{fontenc}

\DeclareRobustCommand{\VAN}[3]{#2}
\let\VANthebibliography\thebibliography
\def\thebibliography{\DeclareRobustCommand{\VAN}[3]{##3}\VANthebibliography}

\usepackage{graphicx}	
\usepackage{amsmath}	
\usepackage{amssymb}	
\usepackage{multirow}
\usepackage{newtxtext,newtxmath}
\defcitealias{Buzzo_22b}{B22}
\defcitealias{Buzzo_24}{B24}
\defcitealias{Ferre-Mateu_23}{FM23}

\title[The multiple classes of UDGs]{The multiple classes of ultra-diffuse galaxies: Can we tell them apart?\thanks{This paper is dedicated to the memory of Prof. Thomas Harold Jarrett, deceased on 3 July 2024.}}

\author[M. L. Buzzo et al.]{Maria Luisa {Buzzo}$^{1,2,3}$,
Duncan A. Forbes$^{1,3}$,
Thomas H. Jarrett$^{4,5,*}$,
Francine R. Marleau$^{6}$,
\newauthor
Pierre-Alain Duc$^{7}$,
Jean P. Brodie$^{1,3,8}$,
Aaron J. Romanowsky$^{9,10}$,
Anna {Ferré-Mateu}$^{11,12,1}$,
\newauthor 
Michael Hilker$^{2}$,
Jonah S. Gannon$^{1,3}$,
Joel Pfeffer$^{1,3}$, and
Lydia Haacke$^{1,3}$
\\ \\
$^{1}$ Centre for Astrophysics and Supercomputing, Swinburne University, John Street, Hawthorn VIC 3122, Australia \\
$^{2}$ European Southern Observatory, Karl-Schwarzschild-Strasse 2, 85748 Garching bei M\"unchen, Germany \\
$^{3}$ ARC Centre of Excellence for All Sky Astrophysics in 3 Dimensions (ASTRO 3D), Australia \\
$^{4}$ Department of Physics and Astronomy, University of the Western Cape, Robert Sobukwe Road, Cape Town, 7535, South Africa \\
$^{5}$ NASA Infrared Telescope Facility, 640 North Aohoku Place, Hilo, HI 96720, USA \\
$^{6}$ Institut für Astro- und Teilchenphysik, Universität Innsbruck, Technikerstraße 25/8, Innsbruck, A-6020, Austria \\
$^{7}$ Observatoire Astronomique de Strasbourg (ObAS), Universite de Strasbourg - CNRS, UMR 7550 Strasbourg, France \\
$^{8}$ University of California Observatories, 1156 High Street, Santa Cruz, CA 95064, USA \\
$^{9}$ Department of Physics and Astronomy, San José State University, One Washington Square, San Jose, CA 95192, USA \\
$^{10}$ Department of Astronomy \& Astrophysics, University of California Santa Cruz, 1156 High Street, Santa Cruz, CA 95064, USA \\
$^{11}$ Instituto Astrofisica de Canarias, Av. Via Lactea s/n, E38205 La Laguna, Spain \\
$^{12}$ Departamento de Astrofisica, Universidad de La Laguna, E-38200, La Laguna, Tenerife, Spain \\
}

\date{Accepted XXX. Received YYY; in original form ZZZ}

\pubyear{2024}

\begin{document}
\label{firstpage}
\pagerange{\pageref{firstpage}--\pageref{lastpage}}
\maketitle

\begin{abstract}
This study compiles stellar populations and internal properties of ultra-diffuse galaxies (UDGs) to highlight correlations with their local environment, globular cluster (GC) richness, and star formation histories. Complementing our sample of 88 UDGs, we include 36 low-surface brightness dwarf galaxies with UDG-like properties, referred to as NUDGes (nearly-UDGs). All galaxies were studied using the same spectral energy distribution fitting methodology to explore what sets UDGs apart from other galaxies. We show that NUDGes are similar to UDGs in all properties except for being, by definition, smaller and having higher surface brightness. We find that UDGs and NUDGes show similar behaviours in their GC populations, with the most metal-poor galaxies hosting consistently more GCs on average. This suggests that GC content may provide an effective way to distinguish extreme galaxies within the low surface brightness regime alongside traditional parameters like size and surface brightness. We confirm previous results using clustering algorithms that UDGs split into two main classes, which might be associated with the formation pathways of a puffy dwarf and a failed galaxy. The clustering applied to the UDGs+NUDGes dataset yields an equivalent result. The difference in mass contained in the GC system suggests that galaxies in different environments have not simply evolved from one another but may have formed through distinct processes.
\end{abstract}




\section{Introduction}
\label{sec:introduction}

Ultra-diffuse galaxies (UDGs) are the topic of extensive debate. Although there are references to extended low surface brightness galaxies dating back to the 1950s \citep{Reaves_56}, it was not until 2015 that they achieved notoriety, when \cite{vanDokkum_15} unexpectedly found dozens of UDGs in the Coma cluster. First thought to mainly populate dense clusters (e.g., \citealt{vanDokkum_15,Yagi_16,Mihos_15,Venhola_17,Venhola_21,Wittmann_17,Gannon_22,Janssens_19,ManceraPina_19, Iodice_23,Marleau_24a}), UDGs are now found in all environments, including the field \citep[e.g., ][]{Yagi_16,MartinezDelgado_16,Barbosa_20,Marleau_21,Zaritsky_19,Zaritsky_21,Zaritsky_23}.
According to the original definition by \cite{vanDokkum_15}, UDGs are galaxies with central surface brightnesses fainter than $\mu_{g,0} = 24$ mag. arcsec$^{-2}$, and with a half-light radius larger than $R_{\rm e} = 1.5$ kpc. This definition has been debated in many works for its arbitrarity and the many selection effects embedded in such criteria. These selection effects are discussed extensively in \cite{vanNest_22} and references therein. In this study, we adopt the standard \cite{vanDokkum_15} definition, bearing in mind the caveats mentioned above that come along with this assumption. We discuss, nonetheless, some of its implications in Section \ref{sec:udg_definition}.

UDGs first received attention for their large numbers in the Coma cluster and their extended sizes; however, as the years passed, more and more unusual properties of UDGs were found. Some interesting properties include having disproportionately large numbers of globular clusters (GCs) for their stellar masses \citep[see e.g.,][]{Forbes_20a}, in some rare cases hosting unusually overluminous GCs \citep{vanDokkum_18,vanDokkum_19b}, unusual dark matter (DM) content \citep[both DM-dominated and DM-depleted, e.g.,][Tang et al. subm, Buzzo et al. in prep.]{vanDokkum_18,vanDokkum_19b,Shen_21,Danieli_19,ManceraPina_19,ManceraPina_19b,ManceraPina_22,Forbes_Gannon_24,Romanowsky_24}, amongst others. More recently, some of these unusual properties have also been found in other low-surface brightness (LSB) galaxies that do not meet the UDG criteria \citep[e.g.,][]{Toloba_23,Forbes_Gannon_24,Gannon_24}.

Some of these unusual properties were shown to correlate. For example, the number of GCs around UDGs is tightly connected to their halo masses. In fact, for galaxies across a wide range of stellar masses, the number of GCs is known to linearly correlate with the halo mass \citep{Spitler_Forbes_09,Harris_13,Burkert_Forbes_20}. This relation was shown to hold for the GC--rich UDG DF44\footnote{We note that DF44 follows the GC--halo mass relation irrespective of if the GC estimate of \cite{vanDokkum_18} of 76 or the estimate of \cite{Saifollahi_22} of 20 GCs is adopted.} in \cite{vanDokkum_19b}, where a direct measurement of the halo mass was obtained. This same halo mass measurement has shown, nonetheless, that this UDG does not follow the standard stellar mass--halo mass relation and instead lives in an overly massive DM halo. Following these findings, many other GC-rich (i.e., with $N_{\rm GC} \geq 20$) UDGs were found to live in overly massive DM halos \citep{Forbes_Gannon_24}. In contrast, GC-poor (i.e., with $N_{\rm GC} < 20$) UDGs were found to follow the stellar mass-halo mass relation as expected \citep{Gannon_22, Toloba_23}. Most UDGs predicted to live in massive halos using their GC numbers are in high-density environments, indicating that the environment likely plays a role in forming such galaxies and shaping their unusual properties. A few cases of group and field GC--rich UDGs are known, for which different formation scenarios likely need to be invoked. The environment has also been shown to influence the GC radial profile of UDGs. In the MATLAS group and field environments, UDGs exhibit GC distributions that closely follow the stellar light, with a typical ratio of $R_{\rm e,GC}/R_{\rm e,gal} \sim 1.0$ \citep{Marleau_24}. However, in cluster environments, the situation varies. While some GC distributions are consistent with those reported by \cite{Marleau_24}, other studies have found more extended GC systems relative to their host galaxies, with $R_{\rm e,GC}/R_{\rm e,gal} \sim 1.5$ \citep[e.g.,][]{vanDokkum_17, Lim_18, Janssens_24}. In contrast, some studies report more concentrated GC distributions \citep[e.g.,][]{Saifollahi_22, Saifollahi_24}.

The environment and dark matter content of UDGs were also shown to correlate with their physical parameters \citep[e.g.,][]{Rong_Yu_20,KadoFong_20,KadoFong_21,vanNest_22,Jiaxuan_23}, especially their luminosities and ellipticities \citep{Rong_Yu_20,Buzzo_24,Pfeffer_24}. \cite{Rong_Yu_20}, for example, showed that there are at least two populations of UDGs, the elongated ($b/a \sim$ 0.4) and the round ones ($b/a \sim$ 0.9), and that the roundest UDGs are brighter than their elongated counterparts. When analysed in terms of clustercentric distance, the UDGs closer to the centre of clusters were found to be the roundest ones and the elongated ones to live in the outskirts or outside the cluster virial radius, bringing back the idea that the environment plays a role in shaping these galaxies. Trends in the axis ratio, mass, and luminosity of UDGs have also been reported in the Coma and Virgo clusters by \cite{Lim_18} and \cite{Lim_20}, respectively. They found that UDGs that are more dark matter dominated (i.e., with total (dark+baryonic matter) mass-to-light ratios greater than 1000) have relatively rounder shapes (higher $b/a$) and have higher GC specific frequencies ($S_N$), while UDGs with lower total mass-to-light ratios of 500 are more elongated (low $b/a$) and have lower $S_N$. A similar trend was recently found using simulations by \cite{Pfeffer_24}.

The stellar populations of UDGs are also expected to exhibit variations based on the number of GCs they host and their environments, as these factors are closely tied to the galaxies' star formation histories. Due to their faintness, obtaining detailed stellar population data for UDGs is challenging, often requiring many hours of exposure to achieve spectra with sufficient signal-to-noise (S/N). Consequently, only a limited number of these galaxies have been thoroughly studied. \cite{Ferre-Mateu_23} conducted the largest spectroscopic study to date, covering 25 UDGs primarily located in clusters. They showed that UDGs scatter around the dwarf mass--metallicity relation \citep[MZR,][]{Simon_19}. Most UDGs were found to be consistent with the classical dwarf MZR, having metallicities as expected for their masses. However, some UDGs were found to be significantly metal-poor, instead following the simulated MZRs of high-redshift galaxies \citep{Ma_15}, suggesting early quenching. Notably, these extremely metal-poor UDGs were all GC-rich. This pattern was also observed through imaging and spectral energy distribution (SED) fitting by \citealt{Buzzo_22b} (hereafter \citetalias{Buzzo_22b}) and \citealt{Buzzo_24} (hereafter \citetalias{Buzzo_24}). 

These trends observed in many UDG properties may be associated with different formation scenarios. Recent studies \citep[\citetalias{Buzzo_22b},][\citetalias{Buzzo_24}]{Ferre-Mateu_23}, for example, have suggested that UDGs that follow the classical dwarf MZR resulted from dwarfs that have undergone some process capable of increasing their sizes, a formation scenario often referred to as `puffy dwarfs'. This `puffing-up' can be caused by, e.g., going through a succession of supernova feedback episodes \citep{diCintio_17} or having high-spin halos \citep{Amorisco_16}. UDGs that are more metal-rich than expected for their stellar masses lie above the classical dwarf MZR and suggest a tidal dwarf-like formation scenario \citep{Haslbauer_19, Duc_14}. On the other hand, the UDGs that lie below the classical dwarf MZR (i.e., more metal-poor than expected for their stellar masses) were suggested to have suffered from early quenching. This formation scenario was explored by \cite{Danieli_22}, suggesting that these types of UDGs may have undergone only the first stages of star formation, including the formation of GCs, then have had their star formation halted, ending up with a stellar body mainly made of disrupted GCs. This hypothesis and others relying on early-quenching are often called `failed galaxy' formation scenarios \citep{vanDokkum_15}.

Individual studies are vital for identifying and highlighting the unique properties of UDGs, as well as for guiding further research. However, to connect and correlate the various observed trends, conducting a statistically significant study of UDGs across diverse environments, GC-richness levels, stellar masses, structural parameters, and stellar populations is crucial. In this work, we address this need by assembling one of the largest photometrically-driven studies of UDGs, comprising 88 galaxies (combining the sample of 29 UDGs from \citetalias{Buzzo_22b} and 59 UDGs in \citetalias{Buzzo_24}, previously studied by \cite{Marleau_21} and followed up with \textit{HST} by \citealt{Marleau_24}) with a wide range of properties to begin mapping correlations and associating trends with formation scenarios. Additionally, we include 36 dwarf galaxies in the MATLAS survey from \cite{Marleau_24}, which have properties similar to UDGs but are slightly outside the \cite{vanDokkum_15} criteria—referred to as NUDGes (nearly-UDGs), a term first proposed by \cite{Forbes_Gannon_24}. These 36 NUDGes serve as a control sample, allowing us to compare them to proper UDGs and assess whether additional properties can be used to select extreme galaxies rather than just size and surface brightness.

This manuscript is structured as follows: in Section \ref{sec:literature}, we describe the UDG samples in \citetalias{Buzzo_22b} and \citetalias{Buzzo_24} used in this study. In section \ref{sec:data}, we describe the NUDGes data, including data collection, GC counts, SED fitting methodology and results. In Section \ref{sec:discussion}, we discuss the collective results, putting together the studies of \citetalias{Buzzo_22b}, \citetalias{Buzzo_24} and NUDGes. In Section \ref{sec:conclusions}, we lay out our conclusions. 
This paper assumes the cosmological parameters from the Planck 2020 collaboration \citep{Planck_20}. Throughout the paper, we use median and median absolute deviation statistics to analyse our data.

\section{Literature Data}
\label{sec:literature}

In this study, we combine the sample of 36 NUDGes \citep{Marleau_24} along with the UDG samples from \citetalias{Buzzo_22b} and \citetalias{Buzzo_24} (first studied by \citealt{Marleau_21}) to test trends found for UDGs and to understand if the same trends exist for other dwarf galaxies that are not within the UDG realm. 

The studies of \citetalias{Buzzo_22b} and \citetalias{Buzzo_24} focused on 29 and 59 UDGs, respectively. The former includes UDGs in different environments, from the field to the massive Coma cluster, but it is heavily dominated by galaxies in clusters. The latter includes only UDGs in low-to-moderate density environments from the MATLAS survey \citep{Marleau_21, Marleau_24}. 
These previous studies provide photometric, structural, and stellar population properties for all 88 UDGs, which is crucial to understanding their differences from the NUDGes.

In what follows, we quickly summarise the data in both studies (i.e., \citetalias{Buzzo_22b} and \citetalias{Buzzo_24}) and the reanalysis that was performed to put these three datasets together.

For all of the galaxies in this study, we estimate their local volume density ($\log \rho_{10}$) as a proxy for the density of the environment where these galaxies reside. 
We use the 2MASS Redshift Survey \citep{Huchra_12} and a K nearest neighbours \citep{Mucherino_09} algorithm to recover $\log \rho_{10}$ for our galaxies. For further details of this calculation, see Appendix \ref{sec:appendix_environment}. 

\begin{figure*}
    \centering
    \includegraphics[width=\textwidth]{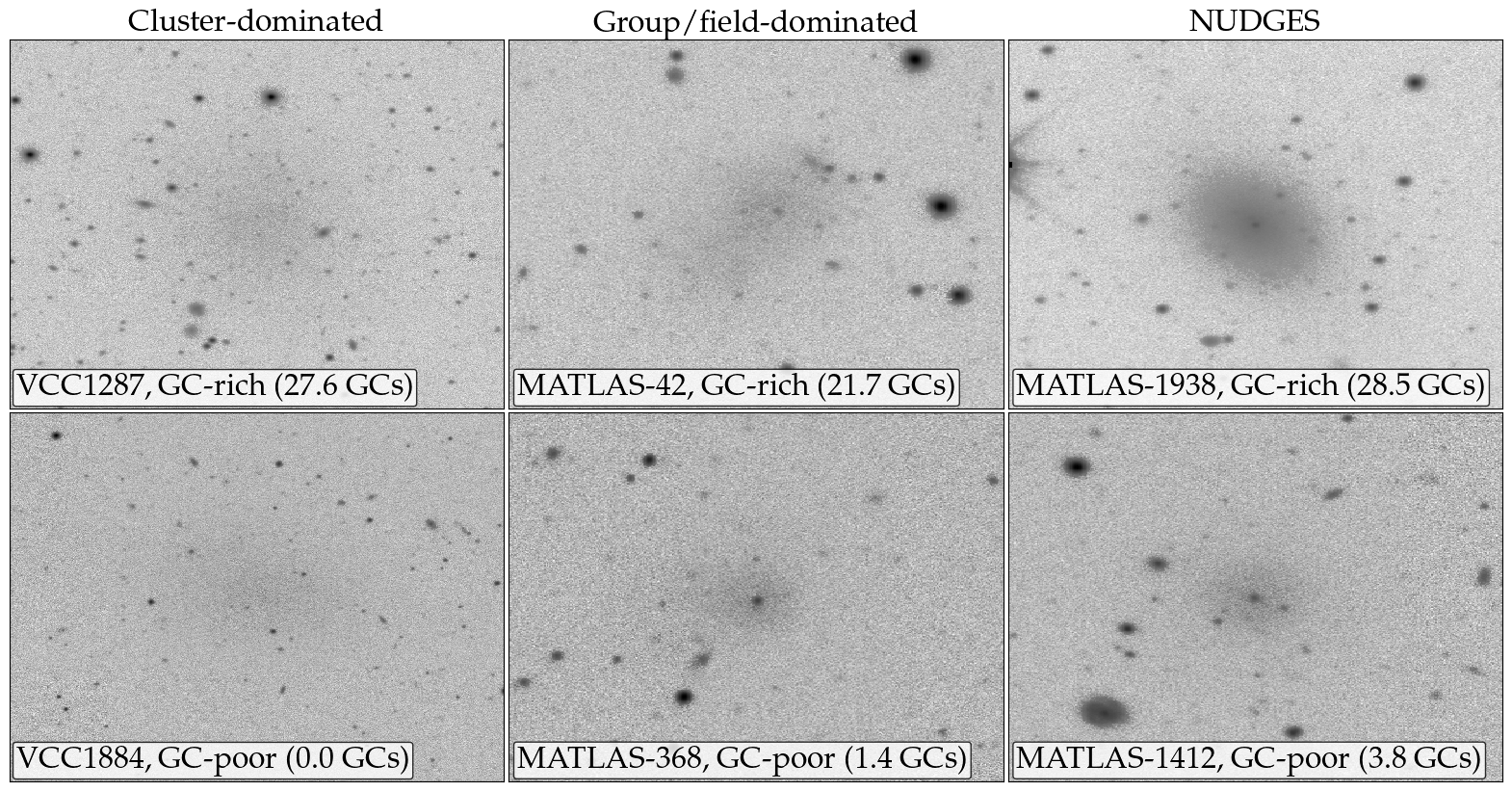}
    \caption{Processed postage stamp $g$-band DECaLS images of galaxies with different GC-richnesses in each of the three samples. The top row shows an example of a GC-rich galaxy, whereas the bottom row shows a GC-poor one. \textit{First column}: Example of UDGs in the cluster-dominated sample (\protect\citetalias{Buzzo_22b}), where GC numbers come from \protect\cite{Lim_20}. \textit{Second column:} UDGs in the group/field-dominated sample (\protect\citealt{Marleau_21,Marleau_24}). The GC numbers come from \protect\cite{Marleau_24}. \textit{Third column:} Examples of NUDGes from \protect\cite{Marleau_24}, where GC numbers come from \protect\cite{Marleau_24}. For all images, the north is up, and the east is left.}
    \label{fig:stamps_three}
\end{figure*}

\subsection{Data from \protect\citetalias{Buzzo_22b} -- Cluster-dominated sample}

The 29 UDGs in \citetalias{Buzzo_22b} include galaxies in the Virgo \citep{Lim_20}, Perseus \citep{Gannon_22} and Coma \citep{vanDokkum_15,Yagi_16,vanDokkum_17} clusters, as well as a few galaxies isolated in the field \citep{Greco_18} and some in groups \citep{Shen_21}. Because of the dominance of galaxies in higher-density environments, we refer to this sample as ``Cluster-dominated'' hereafter. The galaxies in this sample were studied using the Bayesian inference SED fitting code \texttt{PROSPECTOR} \citep{Leja_17,Johnson_21}, using imaging from the optical to the infrared to recover the stellar mass, age, metallicity, star formation timescale and dust attenuation. This work was one of the first to apply \texttt{PROSPECTOR} to such faint galaxies. Although most of the methodology was followed up later in a more detailed study of the 59 UDGs in the MATLAS survey by \citetalias{Buzzo_24}, we realised that some of the configurations and data used in the first study could be improved to recover more reliable stellar populations for the galaxies. Because of that, we refit all of the data from \citetalias{Buzzo_22b} in this study using exactly the same methodology, configuration and dataset as described in \citetalias{Buzzo_24}.

The refitted data have a median stellar mass of $\log (M_{\star}/M_{\odot}) = 8.1 \pm 0.2$, median mass-weighted age of $7.9 \pm 1.3$ Gyr, a population with a median [M/H] = $-1.2 \pm 0.2$ dex, with median $\tau = 1.8 \pm 1.0$ Gyr, where $\tau$ is an approximation of the star formation timescale, i.e., how long does a galaxy take to quench after reaching peak star formation, and small dust content with a median of $A_V = 0.2 \pm 0.2$ mag. These new results are consistent with the median populations of spectroscopically studied UDGs by \cite{Ferre-Mateu_23}, although an offset of $-0.25$ dex is observed in the metallicities. Further details of the reanalysis and results are available in Appendix \ref{sec:revisited_B22}.

The UDGs in this sample span a wide range of GC-richness. The GC-richness classification comes from combining the studies of \cite{vanDokkum_17,Forbes_20a,Lim_20,Gannon_21} and \cite{Saifollahi_22}.

\subsection{Data from \protect\citetalias{Buzzo_24} -- Group/field-dominated sample}
\label{sec:matlasUDGS}

The analysis in the current paper includes the data obtained for the 59 group/field MATLAS UDGs identified by \cite{Marleau_21} and analysed using SED fitting by \citetalias{Buzzo_24}. Because these galaxies are all in groups or isolated in the field, we refer to this sample as ``Group/field-dominated''. For comparison with the other samples of galaxies, the median stellar mass of the MATLAS UDGs is $\log (M_{\star}/M_{\odot}) = 7.6 \pm 0.3$, they were found to have intermediate-to-old ages, with a median mass-weighted age of $7.1 \pm 1.8$ Gyr. They are metal-poor with a median [M/H] of $-1.2 \pm 0.2$ dex. They have a median star formation timescale $\tau$ of $1.6 \pm 0.7$ Gyr and are consistent with no dust attenuation, displaying a median of $A_{V} = 0.12 \pm 0.07$ mag. 

GC numbers were estimated for 38 out of the 59 MATLAS UDGs using single orbit \textit{HST/ACS} data in two filters (F606W and F814W) and are thoroughly described in \cite{Marleau_24}.

\section{New Data}
\label{sec:data}

As previously mentioned, we study 36 dwarf galaxies in the MATLAS survey that are close to the definition of UDGs. Here, we follow the suggestion of \cite{Forbes_Gannon_24} and refer to these galaxies as NUDGes (nearly-UDGs) because they nudge up against the UDG standard definition. This sample of NUDGes was also observed by \cite{Marleau_24} using single orbit \textit{HST/ACS} data for a detailed study of their GC systems. These NUDGes are mainly dwarf elliptical galaxies. Spectroscopic distance measurements of the NUDGes are used when available. When unavailable, we assume that the NUDGes are at the same redshift as the closest massive galaxy to them, following what was suggested by \citetalias{Buzzo_24}. This assumption is mainly based on the recent findings of \cite{Heesters_23}, who used VLT/MUSE to study 56 MATLAS dwarfs and found that 75\% of them were at the same redshift as their hosts. 

Below, we describe the data used to analyse them and the main results. 

\subsection{Imaging}
\label{sec:DECaLS_data}
Archival optical data were obtained for all NUDGes from the Dark Energy Camera Legacy Survey \citep[DECaLS,][]{Dey_19}. None of the galaxies had DECaLS DR10 data available, only DR9, with imaging available in the $g$, $r$ and $z$ bands. For three galaxies, MATLAS-49, MATLAS-203 and MATLAS-207, only images in the $g$ and $r$ bands were available. The reduction and calibration of the DECaLS data are described in \cite{Dey_19}. Although these galaxies were initially characterised (in terms of effective radius and surface brightness) using deep CFHT data \citep{Poulain_21, Marleau_21}, we opt to use DECaLS data in this study. This choice ensures that all galaxies across the three samples are analysed using a consistent dataset for the SED methodology. Moreover, DECaLS provides coverage in at least three bands ($g$, $r$, and $z$) for all galaxies, whereas CFHT data are available in only two bands ($g$ and $r$) for the majority of the galaxies. The classification of the galaxies as UDGs or non-UDGs is maintained from the CFHT determination regardless of the new fitting results obtained with the shallower DECaLS data. 

Following the procedure of \citetalias{Buzzo_24}, we obtain total magnitudes in the optical and structural properties of the galaxies using multi-wavelength galaxy fitting with the \texttt{GALFITM} \citep[][]{Haussler_13, Vika_13} routine. 
Similarly, the process of creation of the PSF, background characterisation and the masking process is described in \citetalias{Buzzo_24}.

A single S\'ersic model was fitted for all galaxies, using the morphological parameters obtained by \cite{Poulain_21} as initial guesses. \texttt{GALFITM} outputs provided the fluxes of the galaxies in each band, as well as their corresponding effective radius ($R_{\rm e}$), S\'ersic index (n), axis ratio ($b/a$) and position angle (PA). By comparing our results, based on DECaLS data, with the measurements from \cite{Poulain_21} for all NUDGes using deeper CFHT data, we find a root mean square (rms) difference of 0.2 mag in the $g$-band magnitude. Similarly, the rms difference for the $g-r$ colour was 0.17 mag, while the difference in central surface brightness was 0.7 mag. Regarding structural properties, the rms differences were 0.32 for the S\'ersic index, 0.12 for the axis ratio ($b/a$), and 1.4 arcsec for the effective radius.

\textit{Wide-field Infrared Survey Explorer} \citep[\textit{WISE}, ][]{Wright_10} imaging was obtained in its four filters (near to mid-infrared), hereafter W1, W2, W3 and W4 for most galaxies. Due to being too faint and small, some galaxies were not detected in \textit{WISE}. Data in the W3 and W4 bands provided upper limits for most galaxies rather than detections. The data are a mix of archival ALLWISE data and bespoke data construction and analysis, including custom mosaic construction from \textit{WISE} single frames. The reduction, calibration and photometric measurement processes are thoroughly described in \citetalias{Buzzo_22b}. We note that the work of \citetalias{Buzzo_22b} included deeper \textit{Spitzer} $3.6$ and $4.5$ $\mu$m imaging, which gave results consistent with those yielded by \textit{WISE} and reinforces that the \textit{WISE} data is suitable for the study of these faint sources. \textit{Spitzer} data is not used in this study for the sake of uniformity with the whole sample.

Optical, near- and mid-IR magnitude measurements are in AB magnitudes and were corrected for Galactic extinction using the two-dimensional dust maps of \citealt{SFD} (recalibrated by \citealt{Schlafly_11}) and the extinction law of \cite{Calzetti_00}.

The photometric measurements in all bands are shown in Appendix \ref{sec:tables}, more specifically in Table \ref{tab:photometry}. Structural parameters obtained from \texttt{GALFITM} are given in Table \ref{tab:morphology}.

\subsection{Globular Cluster Numbers}
\label{sec:gc_counts}
Total globular cluster numbers were obtained for all 36 NUDGes using \textit{HST/ACS} \citep{Marleau_24} (as well as 38 MATLAS UDGs as described in Section \ref{sec:matlasUDGS}). The reduction, source detection, GC candidate selection and final GC counts are thoroughly discussed in \cite{Marleau_24}. GC counts have been used to understand trends of GC--richness with the stellar populations of the galaxies \citep[\citetalias{Buzzo_22b},][\citetalias{Buzzo_24}]{Ferre-Mateu_23}

In Fig. \ref{fig:stamps_three}, we show two examples of UDGs in the cluster-dominated sample (\citetalias{Buzzo_22b}), two UDGs in the group/field dominated sample (\citealt{Marleau_24}, \citetalias{Buzzo_24}), and two NUDGes \citep{Marleau_24}. We choose one GC--rich and one GC--poor galaxy for each sample.

\subsection{SED Fitting}
\label{sec:prospector}

We use the Bayesian Markov Chain Monte Carlo (MCMC) inference code \texttt{PROSPECTOR} \citep[][version 1.2.1]{Leja_17,Johnson_21}, complemented by the Flexible Stellar Population Synthesis package \citep[FSPS;][version 0.4.2]{Conroy_09,Conroy_10a,Conroy_10b}. To sample the posteriors, we use the dynamic nestled sampling \citep{Skilling_04,Higson_19} algorithm \texttt{dynesty} \citep{Speagle_20}. 

The complete description of the configuration and models used in \texttt{PROSPECTOR} is available in \citetalias{Buzzo_22b} and \citetalias{Buzzo_24}.

As mentioned in Section \ref{sec:data}, spectroscopic distance measurements of the NUDGes are used when available. When not available, we assume that the NUDGes are at the same redshift as the closest massive galaxy to them.
We note the caveat that assuming the distance to the closest massive galaxy as the distance to the UDGs and NUDGes may introduce inaccuracies for a significant portion of our sample. However, it is important to mention that many studies in the literature have consistently demonstrated that the distance estimates from the MATLAS team are highly accurate \citep[e.g.,][]{Heesters_23,Buzzo_24,Mueller_24,Kanehisa_24}.

We assume a delayed-$\tau$ exponentially declining star formation history (SFH), as detailed in \citetalias{Buzzo_24}.
The chosen \texttt{PROSPECTOR} configuration has five free parameters: stellar mass (log M$_{\star}$/M$_{\odot}$), metallicity ([M/H]), the onset of star formation (t$_{\rm age}$), star formation timescale ($\tau$) and diffuse interstellar dust attenuation ($A_{V}$). 
We place linearly uniform priors on our free parameters. These are log(M$_{\star}$/M$_{\odot}$) = $6$ -- $10$, [M/H] = $-$2.0 to 0.2 dex, $\tau$ = 0.1--10 Gyr, t$_{\rm age}$ = 0.1--14 Gyr, A$_V = 0 - 4.344$ mag (the full range of the Padova isochrones). Stellar masses are corrected for the currently available mass, rather than the default output representing the total mass ever formed. $t_{\rm age}$ is converted into the mass-weighted age ($t_M$) using a built-in function within \texttt{PROSPECTOR}.

\subsubsection{Median stellar populations of NUDGes}

The best-fit results from \texttt{PROSPECTOR} for the NUDGes are presented in Table \ref{tab:stellarpops}. The median and absolute deviations of the stellar populations of the whole sample of NUDGes are presented below.

We find that the NUDGes have a median stellar mass of $\log (M_{\star}/M_{\odot}) = 7.6 \pm 0.3$, and have old ages with a median mass-weighted age of $t_M = 9.0 \pm 1.4$ Gyr. The galaxies display a median metal-poor population with $[{\rm M/H}] = -1.1 \pm 0.2$ dex. We find a median star formation timescale of $\tau = 0.8 \pm 0.3$ Gyr. Finally, the median internal dust attenuation from the SED fitting of the galaxies in our sample is $A_{V} = 0.16 \pm 0.10$ mag (the data are corrected for Milky Way attenuation). It is interesting to notice that although our dust priors extend out to 4.3 mag, the highest $A_V$ value found is 0.96 mag, highlighting the importance of the inclusion of the \textit{WISE} upper limits from the 12 and 22$\mu$m bands to constrain the amount of dust in the galaxies. This power of the near- and mid-infrared bands in constraining the dust was discussed previously by both \cite{Pandya_18} and \citetalias{Buzzo_22b}.

Individual comparisons between the stellar populations of NUDGes obtained with SED fitting and spectroscopic results from the literature are given in Appendix \ref{sec:comparison_NUDGes}.

\section{Discussion}
\label{sec:discussion}

In this section, we compare UDGs and NUDGes, exploring whether NUDGes resemble specific types of UDGs and examining the potential evolutionary links between these two types of galaxies. We investigate clues that support various formation scenarios and consider whether galaxies with distinct formation histories can transition from one type to another over time.

\subsection{Comparisons between samples}

\begin{figure*}
    \centering
    \includegraphics[width=\textwidth]{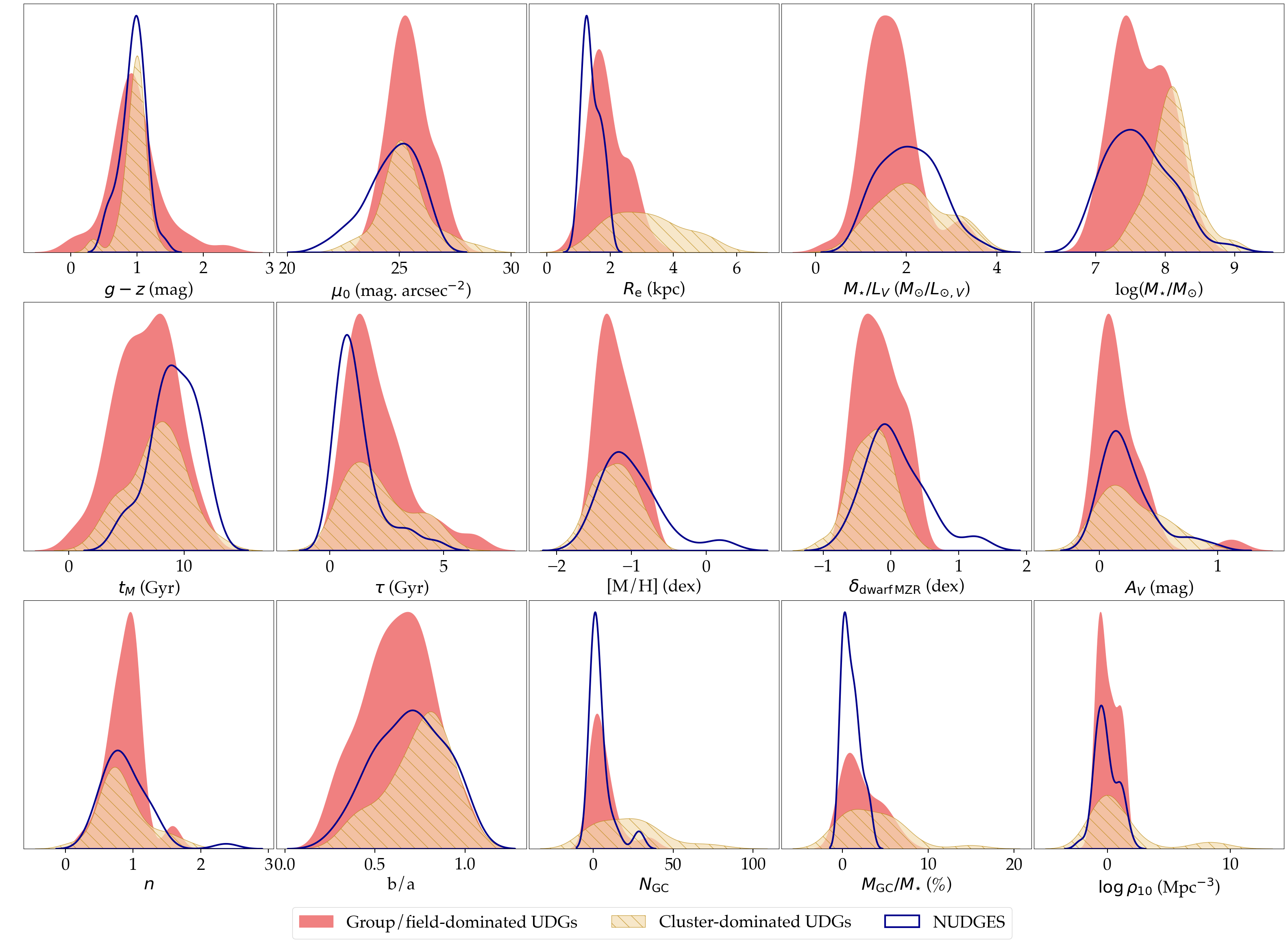}
    \caption{Comparison of the distribution of various properties of the cluster-dominated sample of UDGs, the group/field-dominated sample of UDGs, and NUDGes. The filled red histogram represents the group/field-dominated UDGs, the diagonally-hatched yellow histograms represent the cluster-dominated UDGs, and the blue curves show the distribution of NUDGes. \textit{Left to right, top to bottom}: $g-z$ colour, $g$-band central surface brightness ($\mu_0$), effective radius ($R_{\rm e}$), V-band stellar mass-to-light ratio ($M_{\star}/L_V$), stellar mass (log $M_{\star}/M_{\odot}$), mass-weighted age ($t_M$), star formation timescale ($\tau$), metallicity ([M/H]), distance from the \protect\cite{Simon_19} classical dwarf MZR ($\delta_{\rm dwarf\, MZR}$), dust attenuation ($A_V$), S\'ersic index ($n$), axis ratio ($b/a$), number of GCs ($N_{\rm GC}$), mass of the GC system normalised by the stellar mass of the galaxy ($M_{\rm GC}/M_{\star}$), and local density environment (log $\rho_{10}$). The three distributions exhibit similarities across most properties, with NUDGes generally being smaller, less massive, and slightly brighter than UDGs. The cluster-dominated UDG sample has, on average, higher stellar mass, GC number, and GC system mass than the other two samples.}
    \label{fig:comparison_NUDGes_UDGs}
\end{figure*}

Comparisons between the recovered stellar populations, galaxy structural parameters, environments and GC-richness of the cluster-dominated sample of UDGs, the group/field-dominated sample of UDGs and the NUDGes are shown in Fig. \ref{fig:comparison_NUDGes_UDGs} and further explored in Section \ref{sec:udg_definition}. 
From Fig. \ref{fig:comparison_NUDGes_UDGs}, we observe that while UDGs and NUDGes generally have similar properties, some distinctions suggest that NUDGes may align more closely with either the cluster- or group/field-dominated UDGs in specific traits. NUDGes resemble the group/field-dominated UDGs in aspects such as colour \((g - z)\), surface brightness \((\mu_0)\), local density \((\log \rho_{10})\), and alignment with the MZR. However, they are more similar to the cluster-dominated UDGs in parameters like stellar mass-to-light ratio \((M_{\star}/L_V)\) and star formation timescale \((\tau)\), hinting that some NUDGes may share characteristics with this subgroup.

In terms of stellar mass \((\log M_{\star}/M_{\odot})\), NUDGes fall between the two UDG subgroups, and they show lower relative GC mass fractions \((M_{\text{GC}}/M_{\star})\), differentiating them from both UDG types in this respect. These observations suggest that the NUDGes sample may contain subpopulations with traits that align with either cluster- or group/field-dominated UDGs, depending on the property being considered.

Comparing the UDG samples, we find that cluster-dominated UDGs tend to be more massive \((\log M_{\star}/M_{\odot})\), have higher GC counts \((N_{\text{GC}})\), and show a greater relative GC mass fraction \((M_{\text{GC}}/M_{\star})\) than the group/field-dominated UDGs. These properties are consistent with the expectations for galaxies in denser environments, where higher stellar masses and larger GC systems are more common. Cluster-dominated UDGs also display slightly higher \(M_{\star}/L_V\) and older mass-weighted ages \((t_M)\) than group/field-dominated UDGs. On the other hand, group/field-dominated UDGs generally exhibit a wider range of properties in terms of surface brightness \((\mu_0)\) and star formation timescales \((\tau)\), which may reflect a greater diversity in evolutionary histories within lower-density environments.

Appendix \ref{sec:heatmap} presents and discusses a heatmap that shows the Pearson correlation coefficients between various properties of the combined UDG+NUDGes sample.

\subsection{How different are UDGs and NUDGes?}
\label{sec:udg_definition}

\begin{figure*}
    \centering
    \includegraphics[width=\columnwidth]{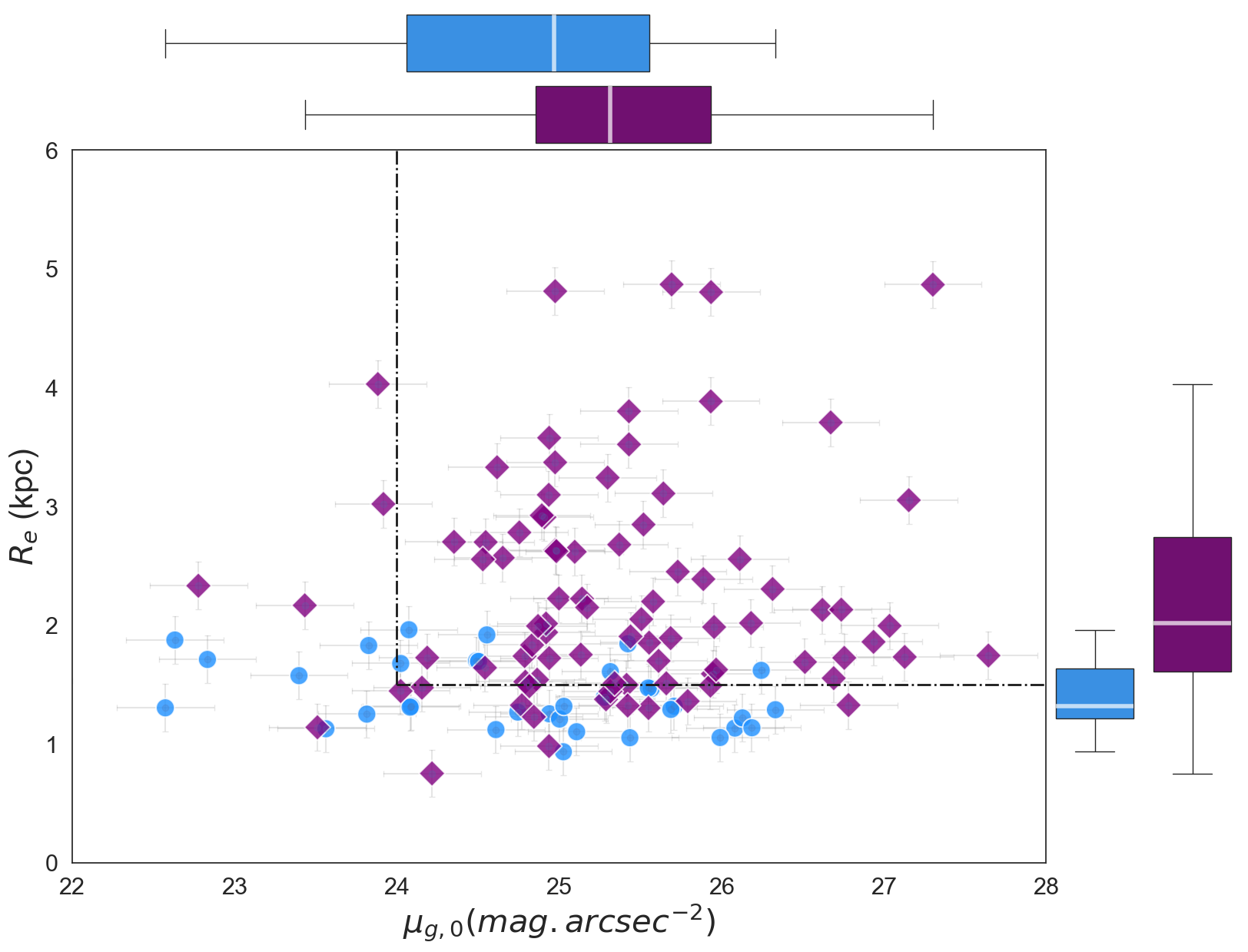}
    \includegraphics[width=\columnwidth]{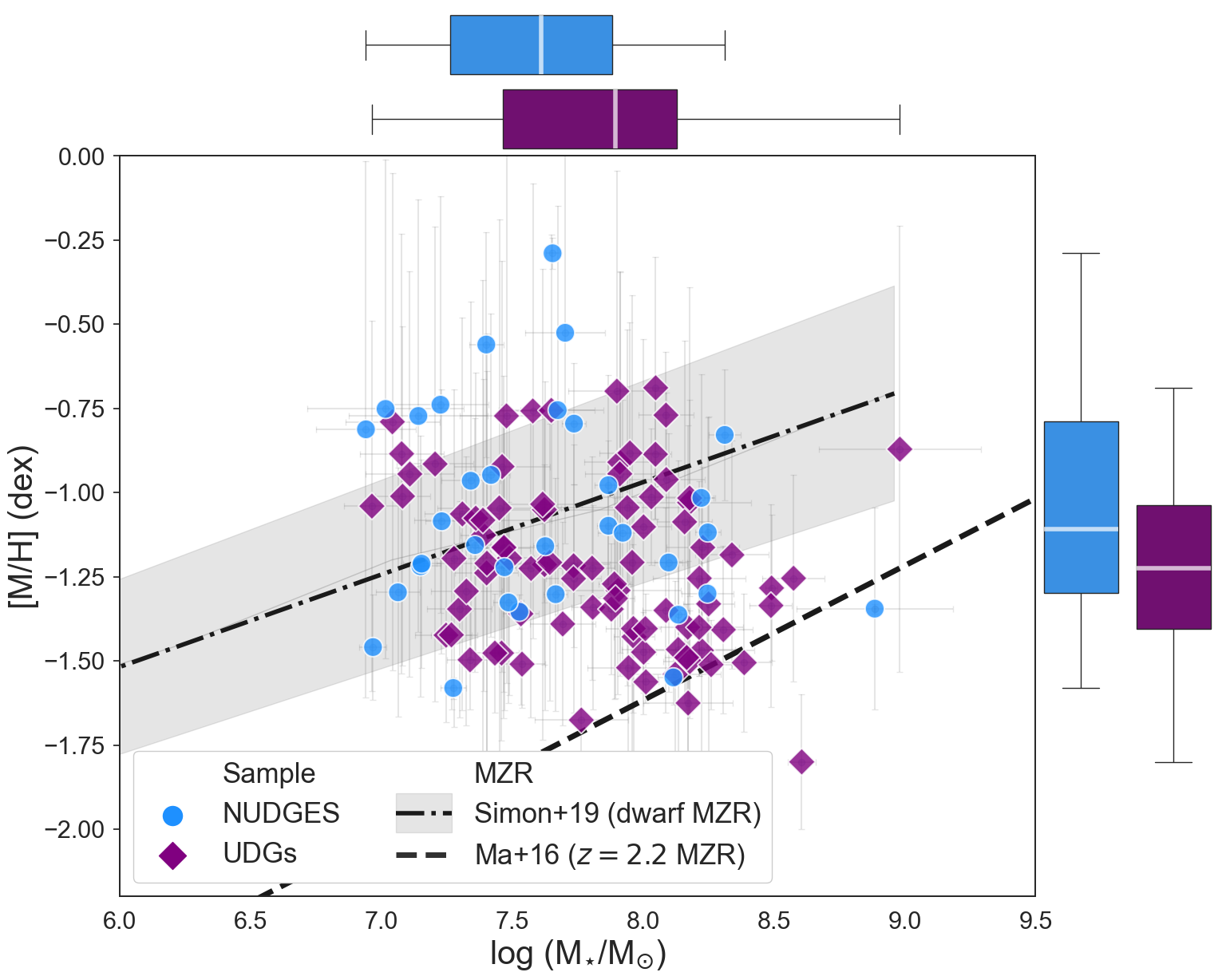}
    \caption{Comparison of UDGs and NUDGes. In both panels, NUDGes \protect\citep{Marleau_24} are the blue circles and UDGs (combination of \protect\citetalias{Buzzo_22b}, \protect\citealt{Marleau_21},\protect\citetalias{Buzzo_24} and \citealt{Marleau_24}) are the purple diamonds. \textit{Left: }Size--luminosity diagram of UDGs and NUDGes. The black dashed lines show the UDG criteria proposed by \protect\cite{vanDokkum_15}. Many NUDGes are within the scatter of the UDG threshold, showing that these galaxies are similar to UDGs. Many UDGs are smaller than the 1.5 kpc effective radius cut due to differences in the galaxy fitting process between this work and the one that first characterised these galaxies as UDGs \protect\citep[i.e., ][]{Poulain_21,Marleau_21}. \textit{Right:} Stellar mass--metallicity distribution of UDGs and NUDGes. The \protect\cite{Simon_19} MZR for classical Local Group dwarf galaxies is shown with the black dash-dotted line, while the grey shading stands for the uncertainty in the relation. The dashed line is the evolving MZR at redshift $z=2.2$ from \protect\cite{Ma_15}. Marginal distributions show the difference between the subsamples. The boxes show the quartiles of the dataset, with the white mark showing the median value, and the black errorbars showing the full extent of the distribution.
    NUDGes are, on average, slightly less massive and more metal-rich than UDGs, but both samples show a similar distribution across the mass-metallicity plane.}
    \label{fig:size_luminosity}
\end{figure*}

The standard classification of UDGs by \cite{vanDokkum_15} has been discussed in many papers for its arbitrarity \citep[e.g.,][]{vanNest_22}. 
In the right-hand side of Fig. \ref{fig:size_luminosity}, we show the size-luminosity distribution of all of the galaxies in this study to try and highlight some of these arbitrarities. In Fig. \ref{fig:size_luminosity}, one can see that many UDGs in our sample are either brighter than the surface brightness threshold or smaller than the effective radius criterion. This is because these UDGs had their properties first determined by \cite{Marleau_21} using deep CFHT data, which yielded properties that fulfilled the UDG criteria. In our study, however, we use DECaLS data, which are slightly shallower and not able to fully detect the total extent of the galaxies (making them smaller than the threshold) or to not accurately measure their true luminosity (making them slightly brighter than the threshold). The same data has also measured some of the NUDGes to be within the UDG region, even though they were found to be smaller and/or brighter in previous works. It is clear, thus, that while the definition of a UDG is not inherently dependent on specific datasets or measurement accuracies, the sample of UDGs identified by this definition is strongly influenced by the data used and the precision with which the galaxies' physical properties are determined. We remind the reader that this study did not utilise the deep CFHT data in order to maintain dataset homogeneity across all galaxies in at least three bands ($g$, $r$, and $z$), as provided by the DECaLS data.

Looking from another perspective, in previous works \citep[e.g., \citetalias{Buzzo_22b},][\citetalias{Buzzo_24}]{Ferre-Mateu_23}, different mass-metallicity relations \citep[i.e.,][]{Kirby_13,Simon_19,Ma_15} were used to show that there are different types of UDGs, some that are similar to classical dwarf galaxies and some that are much more metal-poor than what is expected for their stellar masses, suggestive of a distinct chemical evolution. In the right-hand side of Fig. \ref{fig:size_luminosity}, we investigate whether these observations extend to NUDGes and if these galaxies display similar trends to those observed in UDGs. If so, it might be beneficial to incorporate them into the UDG classification. Including these galaxies could address the surface brightness and size biases inherent to the current UDG definition. Moreover, it would significantly expand the UDG sample, facilitating the compilation of more representative samples of LSB dwarf galaxies for future research.

At first glance, one can see that the NUDGes have higher metallicities and are less massive than UDGs on average. Most of these galaxies seem to follow well the classical dwarf MZR from \cite{Simon_19}. However, some NUDGes were found to lie above the classical dwarf MZR, which can be suggestive of tidal interactions or weak feedback \citep{Collins_Read_22,Sales_22}. On the other hand, five NUDGes were found to lie below the dwarf MZR, with some of them being consistent with the simulated high-redshift MZR from \cite{Ma_15}. 

\begin{figure*}
    \centering
    \includegraphics[width=\textwidth]{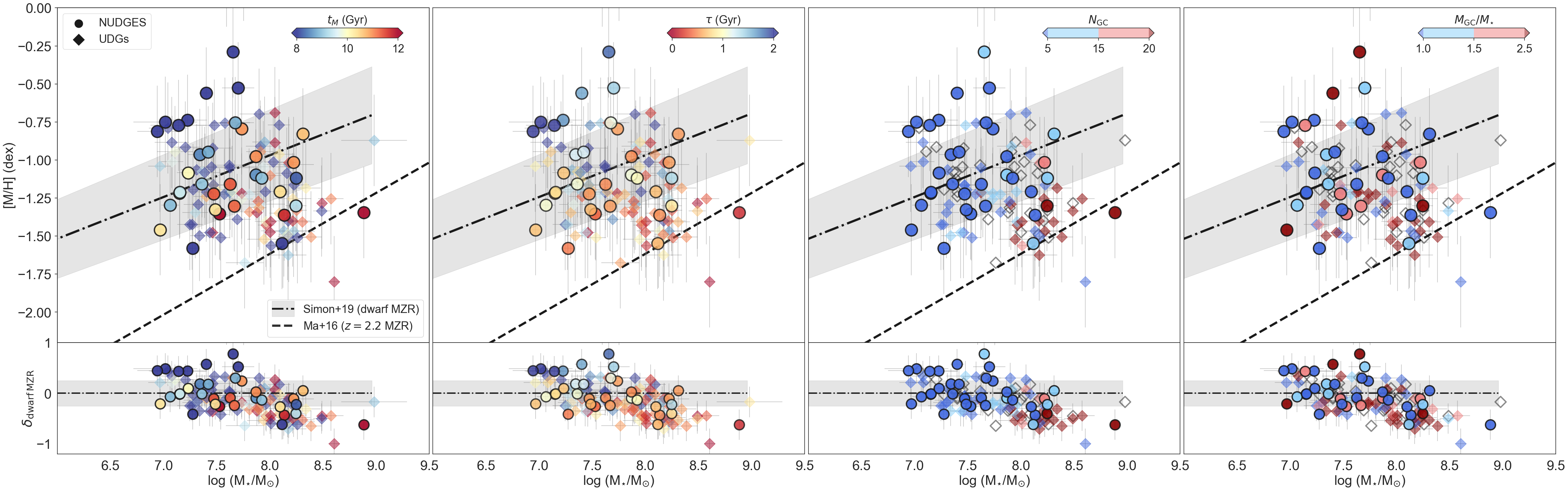}   
    \caption{The stellar mass--metallicity distribution of UDGs (diamonds) and NUDGes (circles). Particular emphasis is placed on the NUDGes by using a higher transparency level for the UDGs. This emphasis is placed because this trend has been previously found for UDGs (\citetalias{Buzzo_24}), and we aim to test if the same trend is found for the NUDGes. The \protect\cite{Simon_19} MZR for classical Local Group dwarf galaxies is shown with the black dash-dotted line, while the grey shading stands for the uncertainty in the relation. The dashed line is the evolving MZR at redshift $z=2.2$ from \protect\cite{Ma_15}. UDGs and NUDGes (\protect\citealt{Marleau_21},\protect\citetalias{Buzzo_22b},\protect\citetalias{Buzzo_24}, \protect\citealt{Marleau_24}) are colour-coded by mass-weighted age ($t_M$), star formation timescale ($\tau$), GC number ($N_{\rm GC}$) and mass of the GC system ($M_{\rm GC}/M_{\star}$), respectively. The unfilled markers in the $N_{\rm GC}$ and $M_{\rm GC}/M_{\star}$ panels are UDGs that lack GC number estimates. The bottom row shows the residual difference between the MZR from \protect\cite{Simon_19} (with the uncertainty in the relation shown in grey) and the metallicity of the galaxies also colour-coded by the same properties. UDGs and NUDGes show similar trends in all properties. The galaxies that follow the classical dwarf MZR are, on average, younger, have prolonged star formation histories, host few GCs and have smaller GC mass over stellar mass. On the other hand, UDGs and NUDGes that lie below the dwarf MZR are older, have shorter star formation histories, and are GC--rich. Interestingly,  the most massive GC systems normalised by stellar mass are also below the MZR, although galaxies with higher GC masses can also be observed within and above the MZR, a behaviour similar to that found with spectroscopy by \protect\cite{Ferre-Mateu_23}.}
    \label{fig:MZR_comparison}
\end{figure*}

Given that the distribution of UDGs and NUDGes in the mass--metallicity plane is similar, in Fig. \ref{fig:MZR_comparison}, we explore if they also show similar trends when we analyse different physical and structural properties. The galaxies are analysed in terms of mass-weighted age, star formation timescale, GC number and GC mass relative to stellar mass. These specific properties were chosen so that the stellar population and GC properties of the two samples of galaxies can be compared.
One can see in Fig. \ref{fig:MZR_comparison} that UDGs and NUDGes that follow the classical dwarf MZR are, on average, younger, have longer star formation histories, are GC--poor, and have only a small GC mass compared to the stellar mass. Conversely, UDGs and NUDGes that are more metal-poor and follow the high-redshift MZR are older, have short star formation histories, host many GCs and have some of the most massive GC systems relative to stellar mass. 
This behaviour is more straightforward to observe in the residual plots shown in the second row of Fig. \ref{fig:MZR_comparison}. This residual makes it easier to quantify how well the galaxies follow the dwarf MZR, and it is defined as the difference between the MZR from \cite{Simon_19} and the metallicity of the galaxies. Based on this parameter, galaxies with $\delta_{\rm dwarf \, MZR} = 0$ perfectly follow the dwarf MZR from \cite{Simon_19}. Galaxies with $\delta_{\rm dwarf \, MZR} > 0$ are above the dwarf MZR, and galaxies with $\delta_{\rm dwarf \, MZR} < 0$ are below the dwarf MZR. 
To quantify the correlation level between $\delta_{\rm dwarf \, MZR}$ and $t_M$, $\tau$, $N_{\rm GC}$ and $M_{\rm GC}/M_{\star}$, we use the Pearson correlation coefficient (more details in Appendix \ref{sec:heatmap}). We find a weak negative correlation with $t_M$ ($-$0.22),  a moderate positive correlation with $\tau$ (0.36), a stronger negative correlation with $N_{\rm GC}$ ($-$0.45) and a moderate negative correlation with $M_{\rm GC}/M_{\star}$ ($-$0.28). These coefficients highlight that the number of GCs and the star formation timescale have the most pronounced influence on $\delta_{\rm dwarf \, MZR}$, and that the mass-weighted age and GC system mass present weaker, but still significant, correlation levels.

We observe a bias in the SED fitting results. GC-rich UDGs (and NUDGes) consistently appear as the most metal-poor galaxies in our sample, consistently below the classical dwarf MZR. However, spectroscopic findings from \cite{Ferre-Mateu_23} indicate that while some GC-rich UDGs do fall below the MZR, others follow it well or are even above it. This discrepancy is partially solved when considering the GC system mass normalised by stellar mass rather than just the number of GCs. When this adjustment is made, UDGs with higher GC system masses are more evenly distributed across the mass-metallicity plane. This suggests that SED fitting might have a stellar mass bias that is corrected when the GC number is normalised by mass, aligning the results more closely with those observed in spectroscopy. One possible interpretation of this, already proposed by \cite{Ferre-Mateu_23}, is that GC-rich UDGs are not always extremely metal-poor, but rather that all extremely metal-poor UDGs, which follow the high-redshift MZR (suggestive of early quenching), are GC-rich. This can be interpreted as evidence that early-quenching scenarios account for both the low metallicity and high GC masses in UDGs (and NUDGes). However, other formation scenarios might also produce systems with massive GC populations but with higher metallicities, such as quenching within dense galaxy clusters.

The UDGs and NUDGes that seem to be the most interesting in our sample, i.e., those that deviate from the canonical dwarf MZR (both above and below it), are the ones with the most significant GC systems, both in terms of numbers and mass.
This seems to suggest that a separation based on properties of the GC system, such as number and/or mass, could provide a useful complement to traditional definitions based on size and luminosity in distinguishing differences between samples of low surface brightness dwarf galaxies.
The analysis suggests that the GC content can set apart extreme low surface brightness galaxies, since this property usually indicates that the galaxies went through significantly different formation histories than regular dwarfs. In the following section, we explore in more detail these observed trends and separations in the properties of both UDGs and NUDGes and the existing correlations between GC properties and the stellar populations of the galaxies.

\subsection{The multiple classes of UDGs (and NUDGes)}
\label{sec:classes}

\begin{figure*}
    \centering
    \begin{subfigure}[t]{0.48\textwidth}
        \centering
        \textbf{Only UDGs} 
        \par\medskip 
        \includegraphics[width=\textwidth,trim={9cm 0 8.5cm 0},clip]{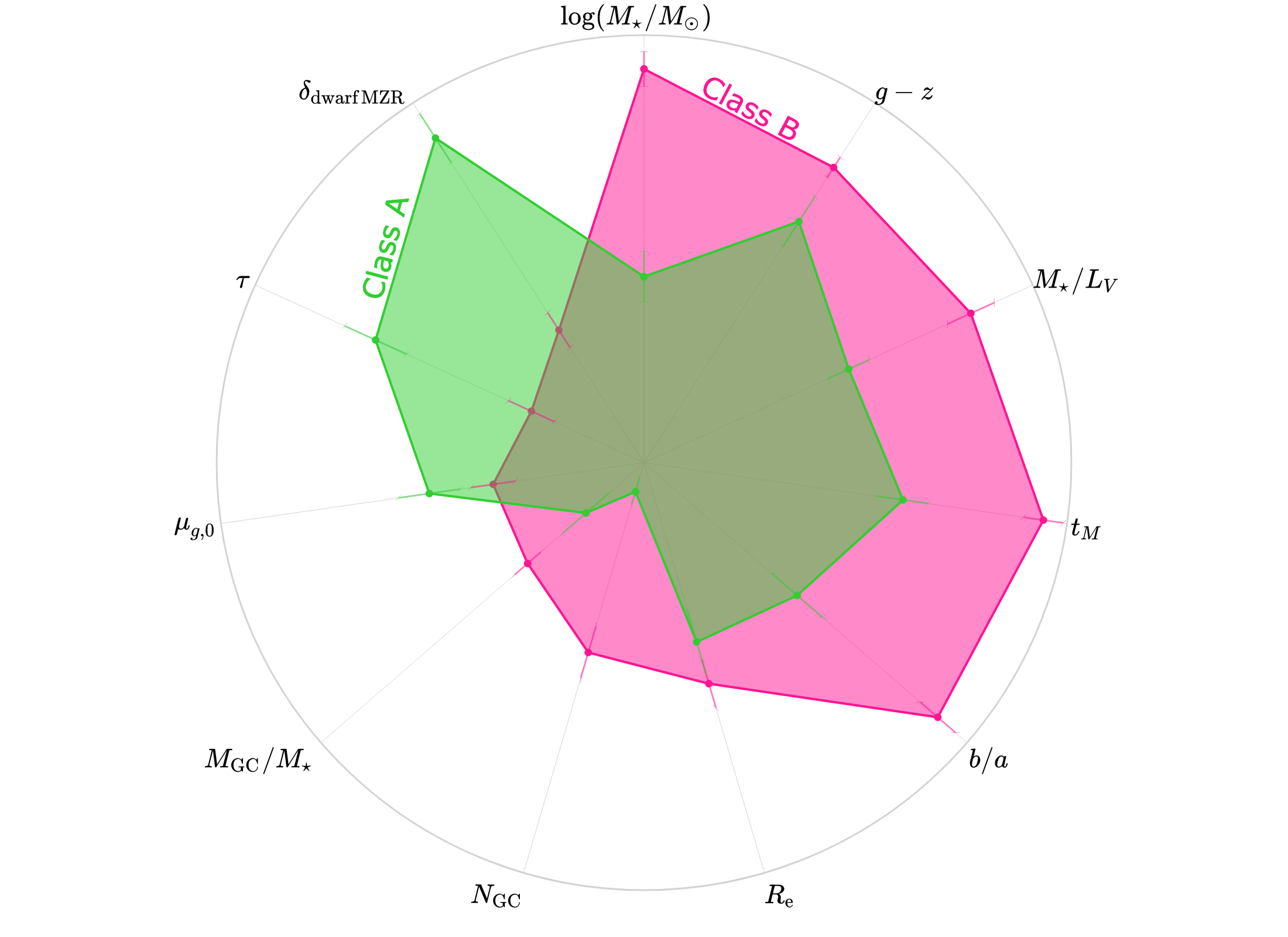}
    \end{subfigure}
    \hfill
    \begin{subfigure}[t]{0.48\textwidth}
        \centering
        \textbf{UDGs + NUDGes} 
        \par\medskip 
        \includegraphics[width=\textwidth,trim={9cm 0 8.5cm 0},clip]{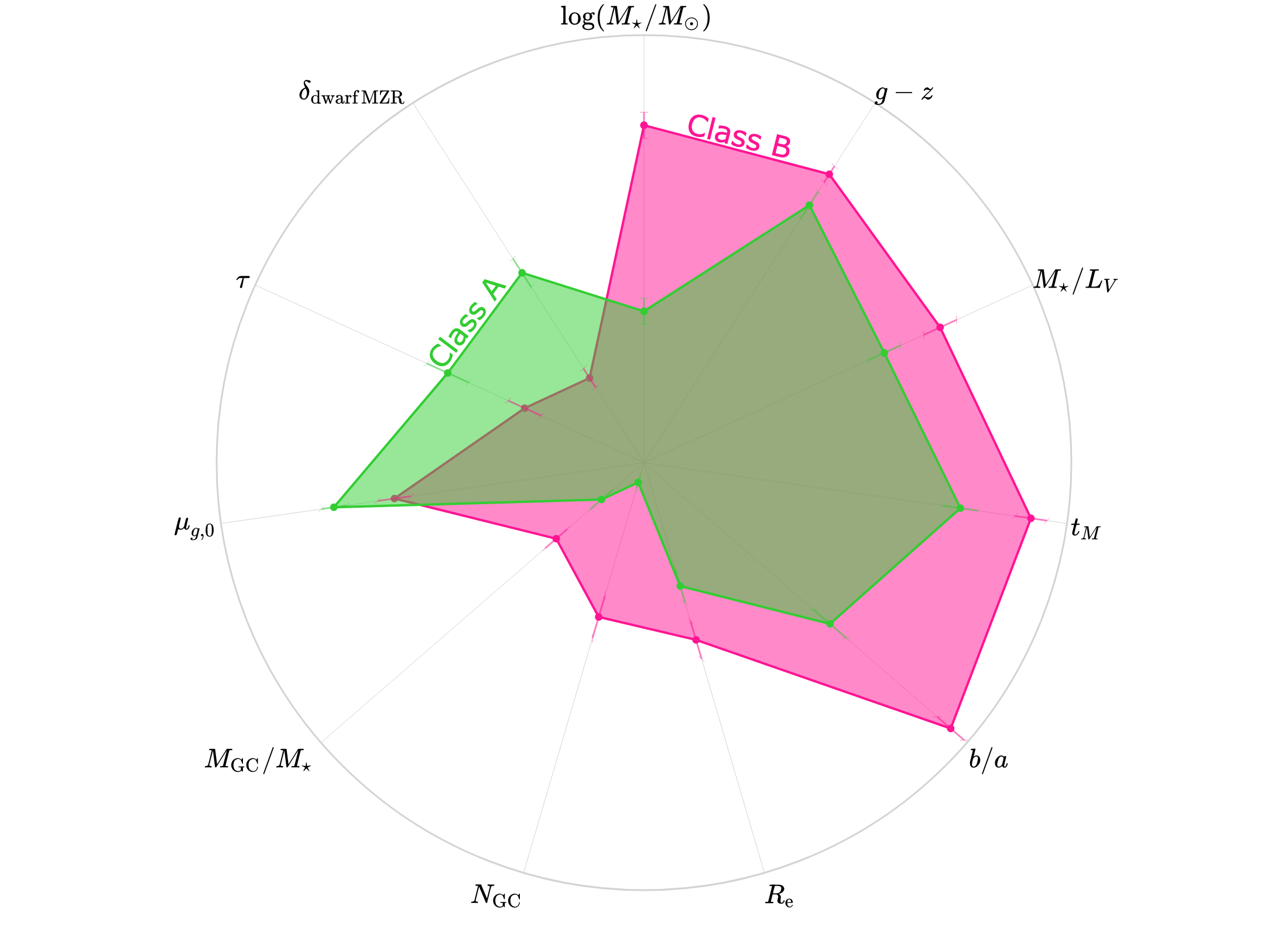}
    \end{subfigure}
    \caption{The results of the \texttt{KMeans} clustering algorithm for the UDG-only and UDG+NUDGes samples. In both panels, the radial axis shows the median value of each property within the classes, while the angular axis represents the properties the clustering algorithm considers. These plots illustrate the relative differences in properties between the galaxies in Classes A and B. Class A is depicted in green and includes UDGs with lower stellar masses, bluer colours, smaller mass-to-light ratios, younger ages (lower $t_M$), more elongated shapes (lower $b/a$), smaller sizes, fewer GCs, lower GC system masses, fainter central surface brightness (higher $\mu_{g,0}$), longer star formation histories (higher $\tau$), and higher $\delta_{\rm dwarf \, MZR}$ (closer to zero, consistent with the classical dwarf MZR). These characteristics align with puffy dwarf-like formation scenarios, suggesting that Class A includes UDGs and NUDGes with this origin. Class B, shown in pink, displays the opposite characteristics: higher stellar masses, redder colours, higher $M_{\star}/L_V$, older ages, rounder shapes (higher $b/a$), larger sizes, more GCs, higher GC system masses, shorter star formation histories (lower $\tau$), and lower $\delta_{\rm dwarf \, MZR}$ (negative values, indicating greater metal deficiency than the classical dwarf MZR). These properties suggest a failed galaxy-like formation scenario for Class B UDGs and NUDGes.}
    \label{fig:polar_two}
\end{figure*}

We begin by applying a clustering algorithm to both the UDG-only sample and the combined UDG+NUDGes sample, following the methodology of \citetalias{Buzzo_24}. This approach allows us to verify if our classifications align with the previous study and assess whether including NUDGes yields consistent results. 

We use the centroid-based clustering algorithm \texttt{KMeans} \citep{Macqueen_67}, an unsupervised machine learning method. \texttt{KMeans} groups galaxies with similar properties in the user-defined multi-parameter space, flagging them accordingly in classes. A detailed description of our implementation of \texttt{KMeans} is available in \citetalias{Buzzo_24}. In this study, the initial parameters for \texttt{KMeans} included the stellar mass ($\log M_{\star}/M_{\odot}$), $g-z$ colour, stellar mass-to-light ratio ($M_{\star}/L_V$), mass-weighted age ($t_M$), axis ratio ($b/a$), effective radius ($R_{\rm e}$), GC number ($N_{\rm GC}$), GC system mass ($M_{\rm GC}/M_{\star}$), central surface brightness ($\mu_{g,0}$), star formation timescale ($\tau$), and the residual metallicity between the galaxies and the classical dwarf MZR ($\delta_{\rm dwarf , MZR}$). Galaxies within the scatter of the classical MZR are assigned \(\delta_{\rm dwarf \, MZR} = 0\). This is done because \texttt{KMeans} does not allow for the incorporation of uncertainties in the clustering analysis, and therefore, we manually adjust \(\delta_{\rm dwarf \, MZR}\) as a pre-processing step to prevent galaxies within the scatter from being misclassified. The criteria for selecting these properties are also discussed in \citetalias{Buzzo_24}. Unlike \citetalias{Buzzo_24}, we exclude the local environment measure from the clustering, as this property is not intrinsic to the galaxies. We later discuss the environmental characteristics of each recovered class. All properties were linearly scaled to a range between 0 and 1 to ensure accurate distance measurements between classes for each parameter analysed.

Similar to \citetalias{Buzzo_24}, we allow \texttt{KMeans} to freely determine the number of classes that best represent the data and use the silhouette score to evaluate the clustering technique. The silhouette score ranges from $-$1 to 1, where $-$1 indicates incorrect class associations, 0 means indistinguishable classes, and 1 signifies perfectly separated classes. Two classes were found to be optimal for both the UDG-only and UDG+NUDGes samples. The UDG-only sample yielded a silhouette score of 0.7, significantly higher than the scores for more than two clusters (i.e., 0.2 was found as the silhouette score for three clusters) and higher than the 0.4 score reported by \citetalias{Buzzo_24}. This suggests that adding the UDGs from \citetalias{Buzzo_22b} to the previous group/field-dominated UDG sample (i.e., \citetalias{Buzzo_24}) strengthens and makes the class separation more robust. The silhouette score for the UDG+NUDGes sample was 0.53, lower than the score obtained using only UDGs, likely due to the broader range of properties—particularly mass-weighted age and star formation timescale—exhibited by the NUDGes.
The polar plots in Fig. \ref{fig:polar_two} show the two UDG classes on the left and the UDG+NUDGes classes on the right. These plots highlight which properties most clearly distinguish the classes and which have less impact. The UDGs associated with each of the classes stayed consistent across both \texttt{KMeans} iterations, i.e., UDGs-only or UDGs+NUDGes.

Focusing first on the UDG-only classification, the two identified classes are summarised in Table \ref{tab:two_types}. The centroids of each property in these classes are consistent with those found by \citetalias{Buzzo_24}, enhancing the reliability and robustness of the median values and identified classes. As in \citetalias{Buzzo_24}, these classes align well with two prominent UDG formation scenarios: Class A resembles classical dwarf galaxies, suggesting a puffy dwarf formation origin, while Class B aligns with early-quenching models, indicative of a failed galaxy formation pathway. Notably, UDGs in Class A are found in less dense environments ($\log \rho_{10}$ = $-0.2$ Mpc$^{-3}$), while those in Class B inhabit denser environments ($\log \rho_{10}$ = 0.2 Mpc$^{-3}$), supporting the hypothesis that the environment plays a role in UDG formation.

The clustering analysis for the UDG+NUDGes sample reveals similar trends to the UDG-only sample, though with less distinct class separation, as reflected by the lower silhouette score. In this classification, 31 NUDGes fall into Class A, aligning with puffy dwarf galaxy formation scenarios, while five are in Class B (the same five mentioned previously to lie below the dwarf MZR), corresponding to failed galaxy scenarios. The final classifications are shown in Fig. \ref{fig:polar_two} and in Table \ref{tab:two_types}. Among the various features considered, the offset from the MZR ($\delta_{\rm dwarf \, MZR}$), the number of GCs ($N_{\rm GC}$) and the axis ratio ($b/a$) emerged as key factors supporting the classification.

For both samples, a slight separation is observed in terms of $M_{\star}/L_V$, as shown in Table \ref{tab:two_types}, which is expected for such metal-poor dwarf galaxies. \cite{Mieske_08} demonstrated that dwarf galaxies with subsolar metallicities have a reasonably constant $M_{\star}/L_V \sim 2$, while those with solar or suprasolar metallicities exhibit higher $M_{\star}/L_V$. Additionally, \cite{Mieske_08} showed that $M_{\star}/L_V$ could be slightly lower than 2, depending on the galaxy’s age, reaching about 1.5 for galaxies around 7-8 Gyr old and closer to unity for galaxies younger than 5 Gyr. We also check that the mean $g-z$ colour obtained for each class aligns with expectations from stellar population synthesis models given these classes’ mean age and metallicity. Finding this separation in an observed property, such as colour, in addition to derived properties like stellar populations, strengthens the results and highlights the intrinsic differences between UDG classes. Examples of galaxies belonging to Class A and B are given in Fig. \ref{fig:stamps_three} (i.e., all three examples in the row of GC--poor galaxies were found to be in Class A, while the row of GC--rich ones contains galaxies in Class B) and the final classification of the galaxies is given in Table \ref{tab:morphology}.

As discussed in \citetalias{Buzzo_24}, a subset of our galaxies lack GC estimates and are thus excluded from this analysis. In \citetalias{Buzzo_24}, we applied the clustering algorithm including and excluding GC information: the former was applied in a sample of 38 galaxies and the latter to all 59 MATLAS UDGs. In both cases, the median properties of the recovered classes were consistent within the uncertainties. In the current work, only the same 21 MATLAS UDGs from \citetalias{Buzzo_24} are affected by missing GC data. Consistent with the previous analysis, we ran the clustering algorithm without GC data across the entire sample of 124 galaxies and found results consistent with those reported in Table \ref{tab:two_types}. These findings confirm that excluding or including GC data does not bias the classification, which remains robust regardless of GC information.

While three UDG classes are expected based on the formation channels discussed in Section \ref{sec:introduction} (failed galaxies, puffy dwarfs, and tidal dwarfs), only two classes were identified in this study and in \citetalias{Buzzo_24}. This is likely because the properties of tidal and puffy dwarfs are generally similar—tidal dwarfs are somewhat more metal-rich and younger on average but not different enough to form a distinct class. Furthermore, tidal dwarfs are exceptionally rare in UDG samples; in our study, for example, only two out of 59 MATLAS UDGs have indicators of being tidal dwarfs. This rarity, combined with the subtlety of their distinguishing characteristics, limits our ability to robustly identify them as a separate class, especially given the constraints of small number statistics.

\begin{table*}
    \centering
    \caption{Median values of classes of UDGs only and UDGs+NUDGes obtained with the \texttt{KMeans} clustering algorithm. The values within brackets show the range of the properties. The properties that were included in the clustering are shown in the first block of the table. The properties not included are in the second block (i.e., divided by a line).}
    \begin{tabular}{c cc cc} \hline
        \multirow{2}{*}{\textbf{Parameter}} & \multicolumn{2}{c}{Only UDGs} & \multicolumn{2}{c}{UDGs + NUDGes}  \\ 
        \cmidrule(lr){2-3} \cmidrule(lr){4-5}
         & \textbf{Class A} & \textbf{Class B} & \textbf{Class A} & \textbf{Class B} \\ \hline
        $\log (M_{\star}/M_{\odot})$ & $7.5 \pm 0.2$ [$7.0 - 8.1$] & $8.1 \pm 0.2$ [$7.5 - 8.6$] & $7.5 \pm 0.1$ [$6.9 - 8.1$] & $8.1 \pm 0.1$ [$7.5 - 8.9$] \\
        $g-z$ (mag) & $0.88 \pm 0.10$ [$0.11 - 1.54$] & $1.02 \pm 0.08$ [$0.60 - 1.28$] & $0.93 \pm 0.04$ [$0.11 - 1.54$] & $1.00 \pm 0.03$ [$0.60 - 1.28$] \\ 
        $M_{\star}/L_V$ & $1.4 \pm 0.2$ [$0.3 - 2.4$] & $2.1 \pm 0.2$ [$1.1 - 3.4$]& $1.8 \pm 0.1$ [$0.3 - 3.6$] & $2.1 \pm 0.1$ [$1.1 - 3.3$] \\ 
        $t_M$ (Gyr) & $5.2 \pm 0.5$ [$0.4  - 10.9$] & $8.4 \pm 0.4$ [$3.7 - 11.2$] & $7.6 \pm 0.4$ [$0.4 - 11.4$] & $9.1 \pm 0.4$ [$3.7 - 12.6$] \\
        $b/a$ & $0.50 \pm 0.05$ [$0.28 - 0.84$] & $0.79 \pm 0.05$ [$0.49 - 0.99$] & $0.58 \pm 0.03$ [$0.28 - 0.99$] & $0.80 \pm 0.03$ [$0.55 - 0.99$]\\
        $R_{\rm e}$ (kpc) & $1.9 \pm 0.2$ [$1.1 - 4.9$] & $2.2 \pm 0.3$ [$1.5 - 5.3$] & $1.4 \pm 0.1$ [$0.8 - 4.9$] & $1.9 \pm 0.2$ [$1.1 - 5.3$] \\ 
        $N_{\rm GC}$ &  $1.3 \pm 0.8$ [$0.0 - 31.4$] & $25.0 \pm 3.7$ [$1.9 - 76.0$] & $1.2 \pm 0.8$ [$0.0 - 13.4$] & $18.6 \pm 3.0$ [$0.0 - 76.0$]\\
        $M_{\rm GC}/M_{\star}$ (\%) & $0.8 \pm 0.4$ [$0.0 - 14.9$] & $3.4 \pm 0.5$ [$0.5 - 8.5$] & $0.7 \pm 0.4$ [$0.00 - 14.9$] & $2.5 \pm 0.4$ [$0.0 - 8.5$] \\ 
        $\mu_{g,0}$ (mag/arcsec$^2$) & $25.4 \pm 0.3$ [$24.2 - 27.6$] & $25.1 \pm 0.2$ [$24.2 - 28.4$] & $25.4 \pm 0.2$ [$22.6 - 27.6$] & $24.9 \pm 0.2$ [$21.7 - 28.4$]\\ 
        $\tau$ (Gyr) & $2.7 \pm 0.3$ [$0.8 - 4.9$] & $1.3 \pm 0.3$ [$0.6 - 4.4$] & $1.7 \pm 0.2$ [$0.4 - 4.9$] & $1.0 \pm 0.2$ [$0.1 - 4.4$]\\ 
        $\delta_{\rm dwarf \, MZR}$ (dex) & $0.02 \pm 0.09$ [$-0.41 - 0.31$] & $-0.46 \pm 0.10$ [$-0.70 - 0.06$] & $0.01 \pm 0.06$ [$-0.13 - 1.26$] & $-0.40 \pm 0.04$ [$-0.70 - 0.24$]\\ \hline 
        [M/H] (dex) & $-1.1 \pm 0.2$ [$-1.5 - -0.7$] & $-1.4 \pm 0.2$ [$-1.6 - -0.9$] & $-1.1 \pm 0.1$ [$-1.6 - 0.2$] & $-1.3 \pm 0.2$ [$-1.6 - -0.8$]\\
        $A_V$ (mag) & $0.2 \pm 0.1$ [$0.0 - 1.0$] & $0.1 \pm 0.1$ [$0.0 - 0.7$] & $0.2 \pm 0.1$ [$0.0 - 1.0$] & $0.1 \pm 0.1$ [$0.0 - 0.7$]\\
        $\log \rho_{10}$ (Mpc$^{-3}$) & $-0.2 \pm 0.1$ [$-0.7 - 1.5$] & $0.2 \pm 0.1$ [$-0.9 - 2.5$] & $-0.3 \pm 0.1$ [$-2.2 - 1.5$] & $0.1 \pm 0.2$ [$-1.0 - 2.5$]\\ \hline 
    \end{tabular}
    \label{tab:two_types}
\end{table*}

In addition to our clustering results, we conduct a preliminary analysis of colour gradients in the UDG and NUDGes samples as proxies for metallicity gradients. Most galaxies in our sample exhibit flat to rising colour gradients, with only three showing possible signs of a decreasing colour gradient. These findings are consistent with those of \cite{Villaume_22}, \cite{Zhao_24}, \cite{Fielder_24} and Ferr\'e-Mateu et al. (in prep). However, recent simulations by \cite{Benavides_24} suggest that UDGs formed through less energetic processes, such as high-spin halos, typically display declining metallicity gradients. In contrast, more energetic events like outflows or tidal stripping tend to result in flatter gradients.
Although our data are of low S/N and insufficient for a detailed analysis of colour gradients, and given the complexities in converting colour gradients to metallicity gradients, our current results do not fully align with these simulation predictions. These discrepancies may be due to differences in the mass ranges between our sample and the simulations, as well as the possibility that the observed colour gradients are driven by age rather than metallicity. Specifically, the flat-to-rising gradients may be influenced by younger stellar populations. In contrast, metallicity gradients remain subtle, as a decrease in [Fe/H] could be counterbalanced by an increase in alpha elements \citep{Pfeffer_22}. 
The predictions from \cite{Benavides_24} align closely with those of \cite{Cardona-Barrero_23}, who used the NIHAO simulations to suggest that UDGs formed through supernova feedback are likely to exhibit flat-to-negative metallicity profiles. Similarly, \cite{Wright_21}, using the ROMULUS25 simulations, predicted that UDGs formed via high-angular-momentum mergers would show steeper negative colour gradients compared to brighter dwarf galaxies.  While our findings diverge from these simulations, this inconsistency is not definitive. Further research, including deeper observational data and spectroscopic studies on a larger sample of UDGs, is necessary to investigate these metallicity gradients more thoroughly.

\subsection{Can classes evolve into one another?}

\begin{figure}
    \centering
    \includegraphics[width=\columnwidth]{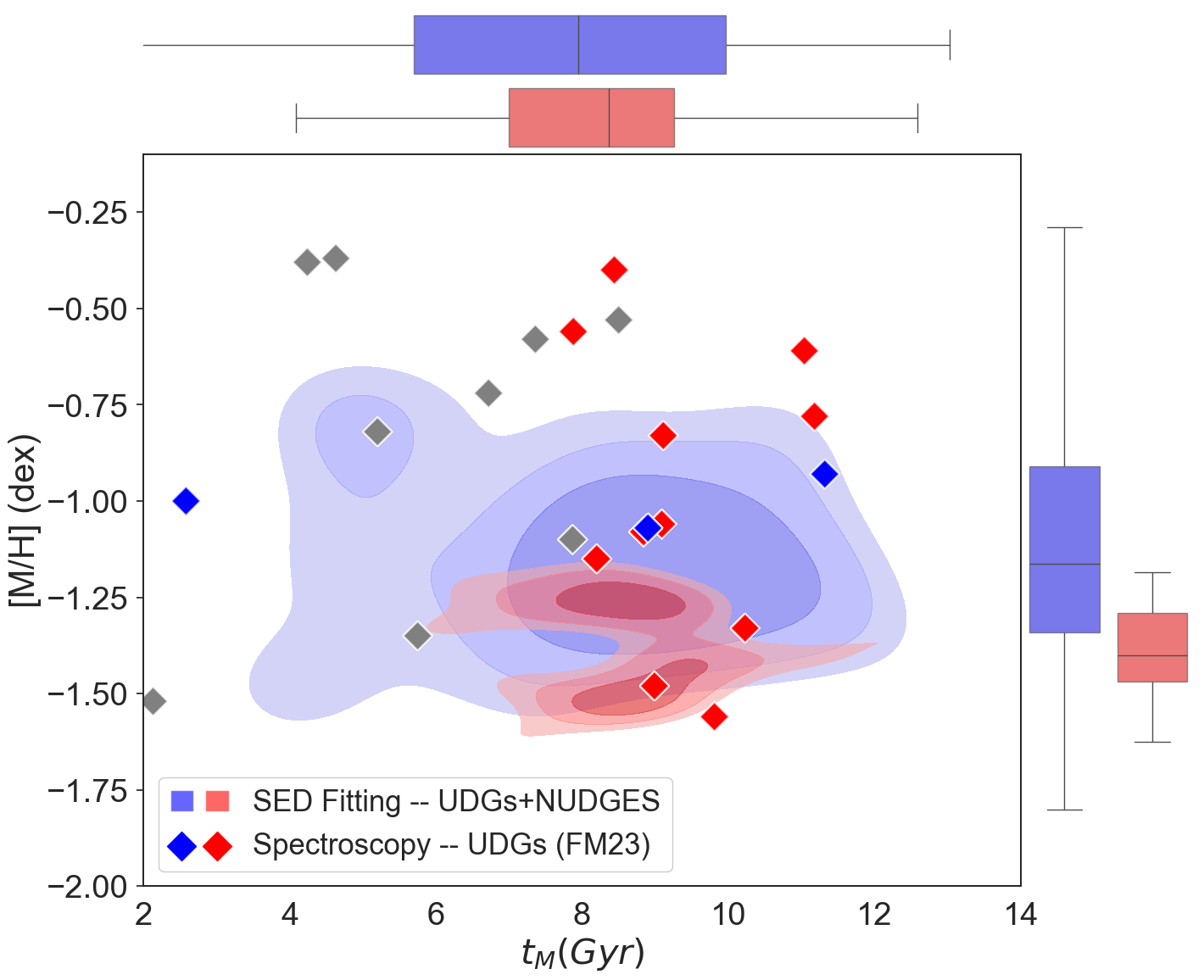}
    \caption{Age-metallicity diagram for all galaxies in this study. GC--rich galaxies are shown in red, while GC-poor ones are in blue. The marginal axes show the 1-d distributions of age and metallicity of the GC-rich and GC-poor samples. The diamonds are UDGs studied with Keck/KCWI spectroscopy by \protect\cite{Ferre-Mateu_23}. GC-rich UDGs are shown in red, GC-poor in blue, and UDGs that lack GC numbers are shown in grey. A separation can be observed in the SED fitting results that is not present in the spectroscopic results (likely due to the bias in the spectroscopic sample towards UDGs in high-density environments): GC--poor UDGs and NUDGes scatter across the age-metallicity plane, while the GC--rich galaxies tend to show consistently lower metallicities.}
    \label{fig:age_met}
\end{figure}

As discussed in the previous section, the properties of the GC systems of low-surface brightness galaxies seem critical to highlight and isolate extreme galaxies from regular dwarfs. This separation in terms of GC content becomes even more evident when comparing the stellar populations of these LSB galaxies (UDGs and NUDGes) according to their GC--richness. This is shown in Fig. \ref{fig:age_met}. The figure illustrates that GC-rich UDGs and NUDGes have different stellar populations than their GC-poor counterparts, with the former being generally older and more metal-poor. A similar trend was previously noted by \citetalias{Buzzo_22b} for a smaller sample of 29 galaxies, and it is encouraging to observe that this pattern holds across a larger sample and is not restricted to UDGs alone. It is worth mentioning, however, that \cite{Ferre-Mateu_23} found through spectroscopy that while GC-rich UDGs tend to scatter around the dwarf mass-metallicity relation (MZR), many exhibit extremely low metallicities, consistent with our findings. While our current study does not replicate the scatter of GC--rich UDGs in the mass--metallicity plane, it does reveal a similar trend when considering the mass of the GC system, with galaxies having a higher percentage of their mass in GCs falling both above and below the MZR. This suggests that systematic effects related to stellar mass, which may be overlooked when considering GC number alone, are revealed when GC number is normalised by mass. This metallicity scatter of spectroscopically studied UDGs can be seen better in Fig. \ref{fig:age_met}.

As discussed earlier and shown in Fig. \ref{fig:polar_two}, our sample of UDGs suggests the presence of at least two distinct populations, with similar trends observed in the NUDGes data. This division raises the intriguing possibility of an evolutionary connection between the two groups. Certain properties, such as axis ratio \citep[where elongated galaxies tend to become rounder through interactions,][]{Moore_96}, age (secular evolution), and metallicity (enrichment from supernovae), can potentially allow galaxies to evolve from one class to another. 
However, while changes in age and metallicity can be individually explained, their combined evolution poses significant challenges. For instance, for such an evolutionary pathway to occur, galaxies would need to either become simultaneously older and more metal-poor, which is highly unlikely, or younger and more metal-rich, which can be explained by new episodes of star formation. Moreover, additional properties—such as the number of GCs and the ratio of GC system mass to stellar mass ($M_{\rm GC}/M_{\star}$)—add further complexity, making an evolutionary connection between these populations difficult to reconcile with existing data.

\begin{figure}
    \centering
    \includegraphics[width=\columnwidth,trim=4cm 2cm 4cm 2cm, clip]{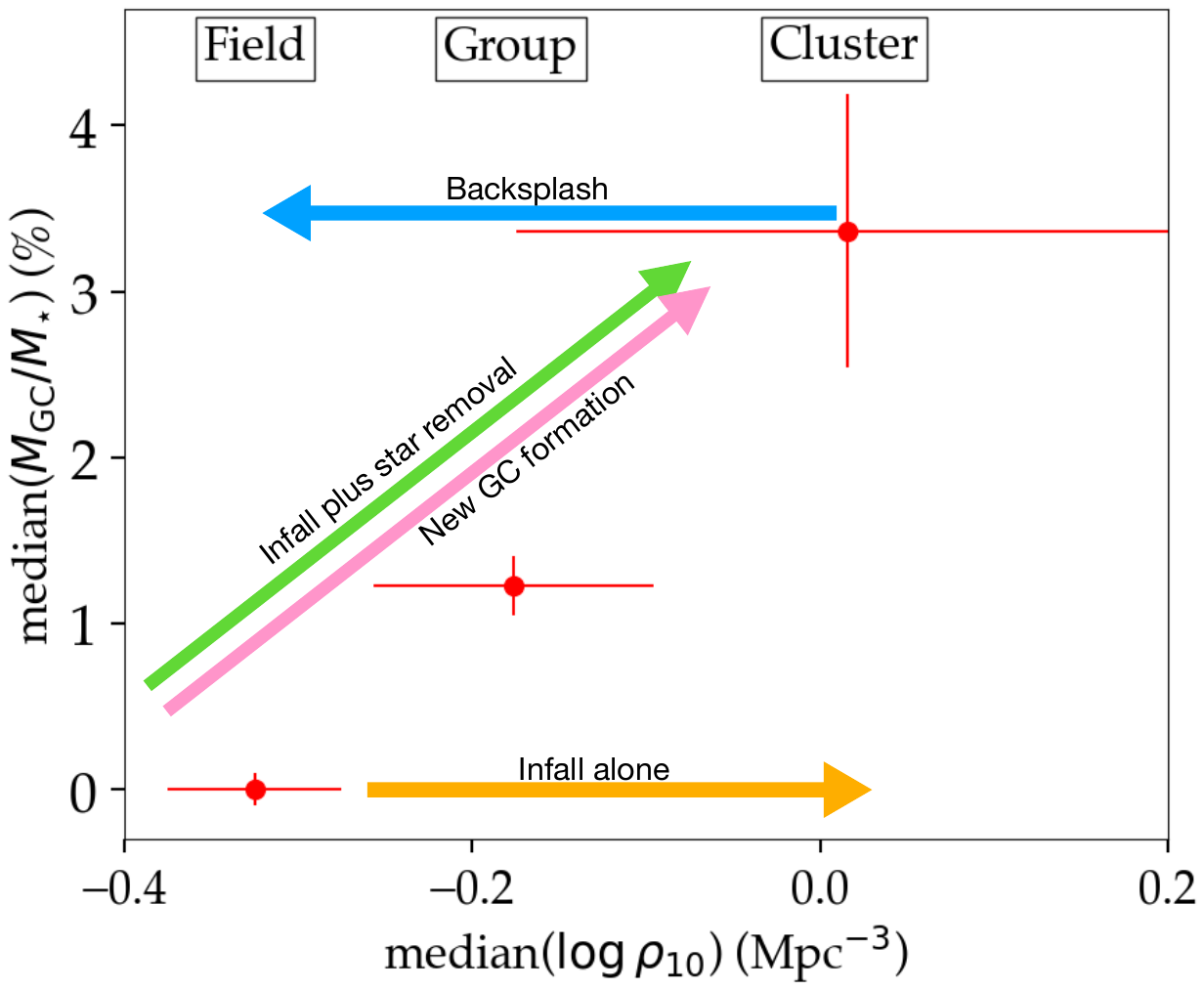}
    \caption{The mass of globular cluster (GC) systems as a function of the local environment for the combined sample of UDGs+NUDGes. The total sample is separated into field (4 galaxies), group (98 galaxies), and cluster (21 galaxies) environments, with the median GC system mass calculated for each environment. The plot demonstrates that galaxies in denser environments, such as groups and clusters, tend to have significantly higher GC system masses than those in the more isolated field environments. The arrows on the plot represent potential evolutionary pathways that galaxies might follow under different formation scenarios. For example, infall alone would not affect the GC system mass. A backsplash scenario \protect\citep{Benavides_21} could potentially return galaxies to the field but does not account for the loss of GC mass. To explain the transition from GC-poor field UDGs+NUDGes to GC-rich cluster UDGs, a combination of infall and star removal or the formation of new GCs would be required. However, these scenarios seem insufficient to account for the large population of GC-rich UDGs observed in clusters. This suggests that galaxies in these different environments have not simply evolved from one another but instead may have formed through distinct processes.}
    \label{fig:NGC_rho10}
\end{figure}

One commonly proposed scenario for explaining the evolution and connection between the two UDG classes is cluster infall (or a similar process in groups or the halos of massive galaxies). 
This is illustrated more clearly in Fig. \ref{fig:NGC_rho10}, which shows the median mass of the GC system divided by the stellar mass of UDGs and NUDGes as a function of their local environment. The classification into field, group, and cluster 
was based on the assumption that all MATLAS UDGs and NUDGes are situated in groups. In contrast, the cluster-dominated sample from \citetalias{Buzzo_22b} was divided based on the candidate host column in Table \ref{tab:morphology}. Applying this separation to the whole sample studied in this work resulted in a distribution of 4 galaxies in the field, 98 in groups, and 21 in clusters. We find that UDGs and NUDGes in the field have a median $M_{\rm GC}/M_{\star}$ of $0.00 \pm 0.10$\% (therefore no GCs) and a median log $\rho_{10}$ of $-0.32 \pm 0.05$ Mpc$^{-3}$. UDGs+NUDGes in groups have $1.23 \pm 0.17$\% and $-0.18 \pm 0.08$ Mpc$^{-3}$, respectively. Finally, UDGs+NUDGes in clusters have $M_{\rm GC}/M_{\star} = 3.36 \pm 0.83$\% and $\rho_{10} = 0.02 \pm 0.19$ Mpc$^{-3}$. The median age and metallicity for the field, group and cluster samples are $t_M = 7.3\, \pm \,0.85$, $7.9\, \pm \,0.3$, $8.2\, \pm \,0.6$ Gyr and [M/H] = $-1.04\, \pm \,0.21$, $-1.20\, \pm \,0.15$, and $-1.25\, \pm \,0.14$ dex, respectively. A clear pattern emerges using these values: isolated galaxies in low-density environments tend to have a smaller fraction of their mass in GCs. In comparison, galaxies in denser environments (such as groups and clusters) have a higher fraction. This suggests that isolated, GC--poor galaxies are unlikely to become GC--rich when they move into denser environments (e.g., via cluster infall). For this to happen, they would need to either form new GCs during infall --an unlikely scenario-- or suffer the combined effect of infall and field star removal in a way that artificially increases the $M_{\rm GC}/M_{\star}$ ratio. However, this explanation does not seem sufficient to explain the large number of GC--rich UDGs observed in clusters. Similarly, GC--rich UDGs in dense environments are unlikely to become GC--poor in isolation, as this would require them to leave the dense environment and to lose GC mass. Although the ``backsplash'' galaxy idea suggested by \cite{Benavides_21} could partially allow for this (i.e., escaping the cluster), it is a rare occurrence and does not account for the many isolated GC-poor UDGs observed. Fig. \ref{fig:NGC_rho10} shows the possible evolutionary paths UDGs (and NUDGes) might take to change environments. The improbability of any single evolutionary trend adequately explaining the large population of UDGs with opposing properties in these distinct environments reinforces the notion that multiple formation scenarios and evolutionary pathways are likely at play, as previously suggested by many works in the literature \citep[e.g.,][]{Lee_17, Papastergis_17,Zaritsky_17,Lim_18,Toloba_18,Prole_19,Jones_23}.

Notably, while $\log \rho_{10}$ measures local rather than global environment, trends observed on larger scales appear to hold at the local level as well: isolated galaxies are predominantly GC-poor, while those in denser environments—whether in groups, the halos of more massive galaxies, or clusters—tend to have more GCs. This effect has also been discussed by \cite{Jones_23}.

We propose that the most distinctive and intriguing galaxies in our sample are those with atypical GC systems relative to their stellar masses, whether in terms of GC numbers or GC system mass. This applies to both UDGs and NUDGes. These galaxies remain challenging for simulations to replicate and lack a fully satisfactory formation scenario, making them prime candidates for future research. Therefore, distinguishing these unusual galaxies based on their GC content, in addition to size and surface brightness, seems like a promising approach for identifying and studying such extreme and enigmatic objects.

\section{Conclusions}
\label{sec:conclusions}
In this study, we use the \texttt{PROSPECTOR} routine to perform spectral energy distribution fitting on 36 NUDGes \citep{Marleau_24}, using data ranging from the optical to the mid-infrared. We extract their structural parameters and stellar populations and compare these to one of the largest photometrically-driven compilations of UDGs (88 in total) analysed using the same methodology in \citetalias{Buzzo_22b} and \citetalias{Buzzo_24}. To ensure consistent analysis and reliable comparisons, we reanalyse the data from \citetalias{Buzzo_22b} using the same methodology applied in \citetalias{Buzzo_24} and for the NUDGes.


Our findings show that many properties of the NUDGes are consistent, within the uncertainties, with those of UDGs. Given these similarities across various properties, we propose that distinguishing extreme galaxies with unusual formation histories—whether UDGs or NUDGes—based on GC system characteristics, such as GC number or GC system mass relative to galaxy stellar mass, could complement traditional criteria based on size and surface brightness.

Following the approach in \citetalias{Buzzo_24}, we apply a clustering algorithm to both the UDGs alone and the combined sample of UDGs and NUDGes. In both cases, the galaxies were divided into two classes. Class A includes the least massive UDGs, characterised by their blue colours, low $M_{\star}/L_V$ ratios, young ages, elongated shapes, small sizes, few or no GCs, minimal GC mass relative to stellar mass, low surface brightness, prolonged star formation timescales, and adherence to the classical dwarf MZR. These properties suggest a formation scenario with dwarf galaxies as progenitors, aligning with puffy dwarf formation models. In contrast, Class B comprises the most massive UDGs, which are redder, have higher $M_{\star}/L_V$ ratios, older ages, rounder shapes, larger sizes, populous GC systems, with a significant fraction of GC mass relative to stellar mass, higher surface brightness, shorter star formation timescales, and deviate from the classical dwarf MZR by lying below it. These characteristics are consistent with a "failed galaxy" scenario, indicative of early quenching. The results were similar for the combined UDGs+NUDGes sample, though clustering was less distinct due to the broader range of ages and metallicities among the NUDGes. The properties that were found to contribute the most to the classification (both to only UDGs and to UDG+NUDGes) were the offset from the MZR ($\delta_{\rm dwarf \, MZR}$), followed by the number of GCs ($N_{\rm GC}$) and axis ratio ($b/a$).

By analysing the fraction of mass in GC systems for UDGs and NUDGes across different environments (field, group, and clusters), we suggest that galaxies with higher GC masses are fundamentally different from those with lower GC masses, implying that these types cannot evolve from one into the other (e.g., by infalling from the field into a cluster).

This work demonstrates that SED fitting is helpful in recovering the collective properties of UDGs and NUDGes and for conducting statistically meaningful comparisons between them.

\section*{Acknowledgements}
This paper is dedicated to the memory of Prof. Thomas Harold Jarrett, whose invaluable contributions and guidance were instrumental in this research. We deeply miss his presence and are profoundly grateful for the time we shared. Tom was essential for developing this work and those that preceded it. Many of the interpretations of this paper came from long discussions with him, where he kept his patience and was always didactic and happy to help. He will be greatly missed by the authors of this manuscript.

This research was supported by the Australian Research Council Centre of Excellence for All Sky Astrophysics in 3 Dimensions (ASTRO 3D), through project number CE170100013. DF, JB and WJC thank the ARC for support via DP220101863. AFM has received support from RYC2021-031099-I and PID2021-123313NA-I00 of MICIN/AEI/10.13039/501100011033/FEDER,UE, NextGenerationEU/PRT. AJR was supported by the National Science Foundation grant AST-2308390.

This paper is based in part on observations from the Legacy Survey, which consists of three individual and complementary projects: the Dark Energy Camera Legacy Survey (DECaLS; Proposal ID \#2014B-0404; PIs: David Schlegel and Arjun Dey), the Beijing-Arizona Sky Survey (BASS; NOAO Prop. ID \#2015A-0801; PIs: Zhou Xu and Xiaohui Fan), and the Mayall z-band Legacy Survey (MzLS; Prop. ID \#2016A-0453; PI: Arjun Dey).
This publication makes use of data products from the Wide-field Infrared Survey Explorer, which is a joint project of the University of California, Los Angeles, and the Jet Propulsion Laboratory/California Institute of Technology, funded by the National Aeronautics and Space Administration.

\textit{Software:} astropy \citep{Astropy_13,Astropy_18}, FSPS \citep{Conroy_10a, Conroy_10b}, python-fsps \citep{Johnson_21a}, \texttt{Prospector} \citep{Leja_17,Johnson_21}.

\section*{Data Availability}
DECaLS data are available via the  \href{https://www.legacysurvey.org/decamls/}{Legacy survey portal}. \textit{WISE} data are available via the \href{https://wise2.ipac.caltech.edu/docs/release/allsky/}{WISE archive}. Globular cluster counts are available in \cite{Marleau_24}. Literature data for the UDGs used for comparison in this study are available in \cite{Buzzo_22b} and \cite{Buzzo_24}.



\bibliographystyle{mnras}
\bibliography{bibli} 




\appendix

\section{Environment measurements}
\label{sec:appendix_environment}
For the NUDGes and the group/field-dominated sample of UDGs, we have an estimate of the local volume density $\log \rho_N$ calculated using the galaxies in the ATLAS$^{\rm 3D}$ survey extending out to 50Mpc. More details about the environment determination can be found in \cite{Duc_14} and \cite{Marleau_21}. Such a measurement does not exist for the galaxies in \citetalias{Buzzo_22b}, and a similar analysis cannot be done for these galaxies as many are at distances greater than 50Mpc, the limit of the ATLAS$^{\rm 3D}$ survey. Because of this, we have re-estimated $\log \rho_{10}$ for all galaxies in our samples, including the NUDGes, group/field-dominated sample of UDGs and cluster-dominated sample of UDGs in \citetalias{Buzzo_22b}. 
We used the 2 MASS Redshift Survey \citep[2MRS,][]{Huchra_12} and a K nearest neighbours \citep{Mucherino_09} algorithm. We have selected only galaxies with magnitudes brighter than 10.5 in the K band in the 2MRS survey. We use the 10 nearest neighbours to map the local volume density of the galaxies. The density was calculated as $\log \rho_N = \frac{3N}{4\pi r_{\rm 3D}}$, where N is the number of neighbours (i.e., 10) and $r_{\rm 3D}$ is the radius at which the most distant neighbour is located. 

Notably, this measurement of the local volume density does not separate central from satellite galaxies; it simply quantifies the local environment that the galaxies reside in. 

In the last panel of Fig. \ref{fig:comparison_NUDGes_UDGs}, we show the distribution of $\log \rho_{10}$ for our three samples of galaxies. One can see that the cluster-dominated sample is the most comprehensive one, ranging from very low to very high-density environments. This sample has the highest average  $\log \rho_{10}$ amongst the three studied. The group/field-dominated sample of UDGs from \citetalias{Buzzo_24} show a bimodality in the density distribution, with one mode being consistent with the average value of \citetalias{Buzzo_22b} and another containing galaxies in lower-density environments. The mode with the more isolated galaxies is similar to that of the NUDGes. Overall, we show that with our three samples of galaxies, we cover a wide range of environments, which will allow us to probe the role of the environment in the evolution and overall properties of the galaxies.

\section{Comparison with the literature}

\subsection{Revisited data from B22}
\label{sec:revisited_B22}

The 29 UDGs in \citetalias{Buzzo_22b} were fitted with the Bayesian inference SED fitting code \texttt{PROSPECTOR}, using imaging from the optical to the infrared to recover the stellar mass, age, metallicity, star formation timescale and dust attenuation of the galaxies. This work was the first to apply \texttt{PROSPECTOR} to such faint galaxies. Although most of the methodology was followed up later on a more detailed study of the 59 UDGs in the MATLAS survey by \citetalias{Buzzo_24}, we realised that some of the configurations and data used in the first study could be improved to recover more reliable stellar populations for the galaxies. Because of that, all of the data in \citetalias{Buzzo_22b} was refitted in this study using exactly the same methodology, configuration and dataset as described in \citetalias{Buzzo_24}.
Some of the reasons why the data were refitted were: (1) the ages reported in \citetalias{Buzzo_22b} were not the mass-weighted ones, but rather the default output ages from \texttt{PROSPECTOR}, i.e., the age since the onset of star formation. (2) aperture photometry was used in \citetalias{Buzzo_22b} instead of total. We found total magnitudes to represent the galaxies' integrated stellar populations better. (3) total stellar masses were reported in \citetalias{Buzzo_22b} instead of mass currently available. (4) DECaLS DR9 data were used instead of DR10, as it was not available then. (5) \texttt{Prospector} was shown to provide consistently older ages in version 1.0 (used in \citetalias{Buzzo_22b}) than version 1.2.1 (used in \citetalias{Buzzo_24}). The former was shown to be less in agreement with spectroscopy. 

\begin{figure}
    \centering
    \includegraphics[width=\columnwidth]{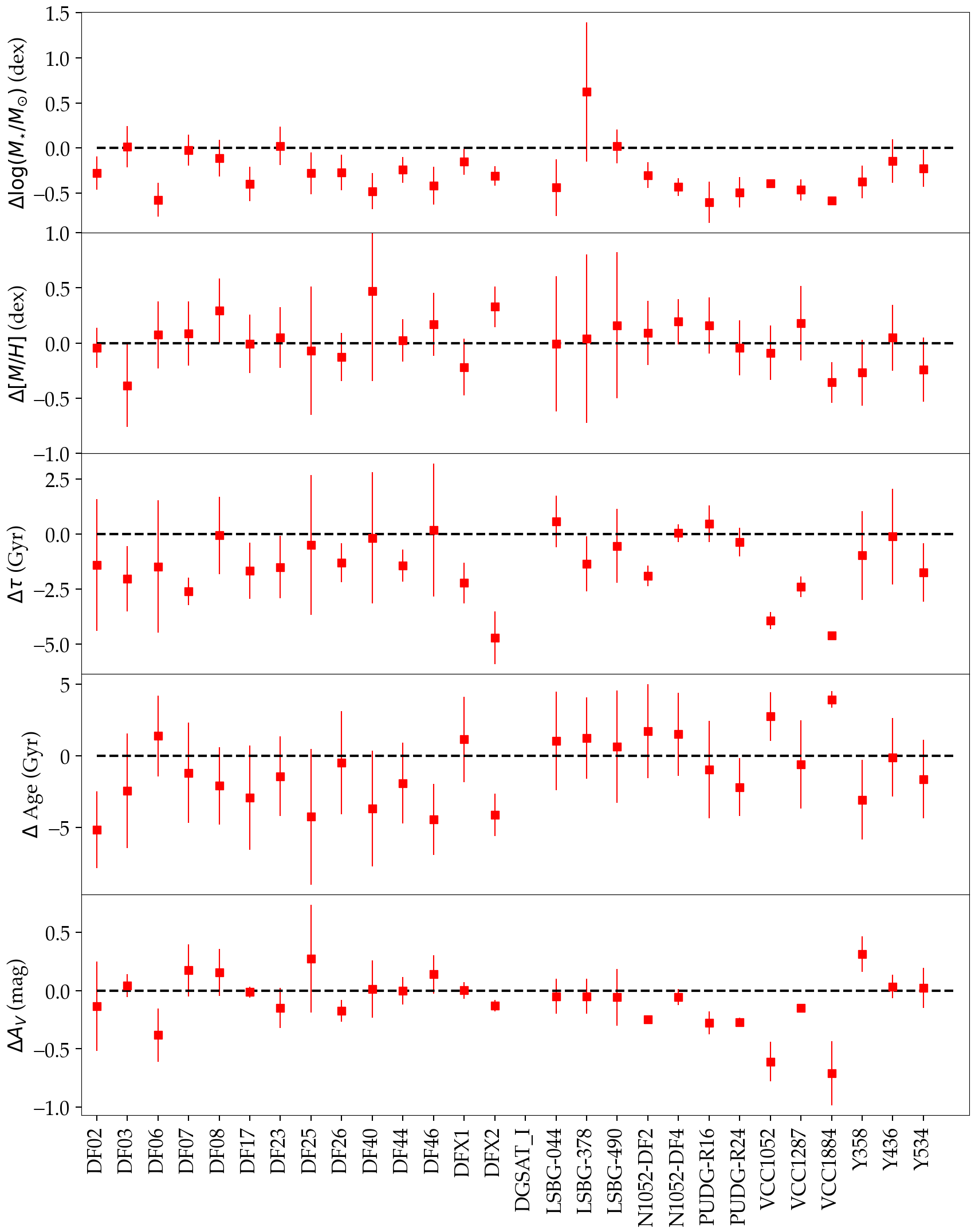}
    \caption{Comparison of the stellar populations of the UDGs obtained with \texttt{PROSPECTOR} in \citetalias{Buzzo_22b} using our previous methodology and the one used in this work. The new methodology provides consistently smaller stellar masses, similar metallicities, younger ages, shorter star formation timescales, and less dust attenuation. These changes are attributed both to the different version of \texttt{PROSPECTOR} used and a different configuration of the routine designed to be more consistent with the type of galaxies we are studying.}
    \label{fig:comp_B22_before_after}
\end{figure}

After refitting the data and correcting for all of the issues mentioned above, we find a median difference in stellar mass between the results of \citetalias{Buzzo_22b} and now of $-0.3$ dex, as the galaxies are consistently less massive now that we quote mass currently available rather than mass ever formed. No significant difference was found in [M/H] (0.02 dex) or dust attenuation ($-0.05$ mag). The new version of \texttt{PROSPECTOR} has delivered consistently younger ages and shorter star formation timescales, with a median difference of $-1.2$ and $-1.4$ Gyr, respectively. In Fig. \ref{fig:comp_B22_before_after}, we show the difference in the recovered properties of the galaxies in \citetalias{Buzzo_22b} and here. 

The refitted data has a median stellar mass of $\log M_{\star}/M_{\odot} = 8.1 \pm 0.2$, a median age of $7.9 \pm 1.3$ Gyr, a population with a median [M/H] = $-1.2 \pm 0.2$ dex, on average short star formation timescales with a median $\tau = 1.8 \pm 1.0$ Gyr, and small dust content with a median of $A_V = 0.2 \pm 0.2$ mag. These new results are consistent with the median populations of spectroscopically studied UDGs by \cite{Ferre-Mateu_23}.

\begin{figure*}
    \centering
    \includegraphics[width=0.8\textwidth]{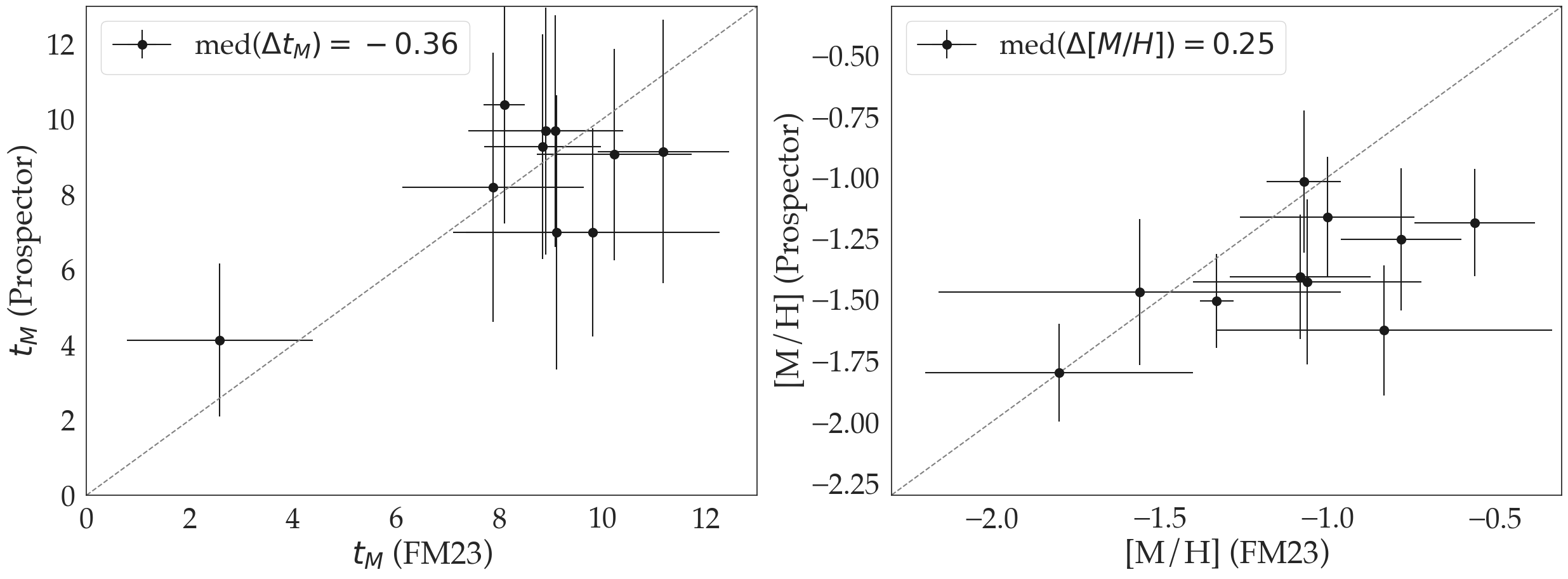}
    \caption{Comparison of the stellar populations of the UDGs obtained with \texttt{PROSPECTOR} and the ones obtained with spectroscopy in \protect\cite{Ferre-Mateu_23}. \textit{Left:} Comparison of recovered mass-weighted age. We find a median difference of -0.36 Gyr. \textit{Right:} Comparison of the recovered metallicity. We find a median difference of 0.25 dex. We attribute this offset partially to the uncertainties in the SED fitting method, but we note that some of the spectroscopic results may be compromised by narrow wavelength ranges, as discussed by \protect\cite{Ferre-Mateu_23}.}
    \label{fig:comp_B22_Anna}
\end{figure*}

In Fig. \ref{fig:comp_B22_Anna}, we compare the age and metallicity results obtained with \texttt{PROSPECTOR} with those obtained from spectroscopy by \cite{Ferre-Mateu_23}. We find a median difference of -0.36 Gyr in age and -0.25 dex in metallicity, which is well within the uncertainties of \texttt{PROSPECTOR}. We note also that some of the spectroscopic results may be compromised by narrow wavelength ranges, as discussed by \cite{Ferre-Mateu_23}. This result emphasizes, nonetheless, that SED fitting is capable of recovering reliable stellar population properties for UDGs.

\subsection{Comparison of the stellar populations of NUDGes as derived by SED fitting and spectroscopy}
\label{sec:comparison_NUDGes}

Three galaxies in our sample of NUDGes were previously studied with spectroscopy by \cite{Heesters_23} using VLT/MUSE data. In this subsection, we compare the results obtained with SED fitting with those obtained spectroscopically.

\begin{itemize}
    \item MATLAS-290

    \cite{Heesters_23} reported a mass-weighted age and metallicity of $t_M = 11.5^{+2.5}_{-0.5}$ Gyr and [M/H] = $-1.39^{+0.01}_{-0.13}$, respectively. They also found a mass-to-light ratio (M/L) of $1.9$ and $g = 18.8$ mag, i.e., $\log_{10}(M_{\star}/M_{\odot}) \sim 8.6$. In this study, we find a mass-weighted age of $t_M = 10.4^{+2.4}_{-3.6}$ Gyr, a metallicity of [M/H] = $-1.1^{+0.4}_{-0.3}$ dex, and a stellar mass of $\log_{10}(M_{\star}/M_{\odot}) = 8.1^{+0.1}_{-0.1}$ for the same galaxy. Our results are consistent with the ones found by \cite{Heesters_23} within 1$\sigma$.

    \item MATLAS-1400

    \cite{Heesters_23} found a mass-weighted age of $t_M = 11.3^{+0.7}_{-2.7}$ Gyr, a metallicity of [M/H] = $-1.20^{+0.07}_{-0.06}$, a M/L of 2.0 and $g = 17.1$ mag, yielding a stellar mass of $\log_{10}(M_{\star}/M_{\odot}) \sim 7.9$. We found results consistent within the uncertainties with \cite{Heesters_23} for the age and stellar mass, with $t_M = 8.0^{+4.1}_{-4.5}$ Gyr, and a stellar mass of $\log_{10}(M_{\star}/M_{\odot}) = 7.7^{+0.2}_{-0.2}$. The metallicity estimated with SED fitting of [M/H] = $-0.43^{+0.63}_{-0.65}$ dex was found to be much higher than the one estimated with spectroscopy, but with large uncertainties. However, as noted in previous works (e.g., \citetalias{Buzzo_24}), the stellar populations in \cite{Heesters_23} appear to be somewhat more metal-poor than expected, especially when compared with findings from other spectroscopic studies (see \citealt{Mueller_20}).
    
    \item MATLAS-1408

    \cite{Heesters_23} found a mass-weighted age of $t_M =  12.9^{+0.9}_{-3.3}$ Gyr, a metallicity of [M/H] = $-1.39^{+0.08}_{-0.20}$, a M/L of 2.1 and $g = 18.8$ mag, i.e., a stellar mass of $\log_{10}(M_{\star}/M_{\odot}) \sim 8.6$. For this galaxy, we have found with \texttt{PROSPECTOR} $t_M = 11.7^{+1.4}_{-2.8}$ Gyr, a metallicity of [M/H] = $-1.3^{+0.2}_{-0.2}$ dex, and a stellar mass of $\log_{10}(M_{\star}/M_{\odot}) = 8.1^{+0.1}_{-0.1}$. Again, our results are consistent within the uncertainties with the literature, although they found a higher stellar mass which can be explained by their deeper CFHT data.
\end{itemize}

Overall, the NUDGes studied by us using \texttt{PROSPECTOR} have yielded stellar populations consistent with those reported from spectroscopy by \cite{Heesters_23}. These comparisons make us confident that within the uncertainties, we are recovering reliable stellar populations for the galaxies in this study, both UDGs and NUDGes.

\section{Tables}
\label{sec:tables}

In this appendix, we provide the photometry, physical properties and stellar population parameters of all of the NUDGes and the 29 refitted UDGs in the cluster-dominated sample from \citetalias{Buzzo_22b}. The properties of the group/field-dominated sample of UDGs used in this study can be found in \citetalias{Buzzo_24}. 

\label{sec:appendix_galfitm_prospector}

\begin{table*}
\scalebox{0.9}{
\begin{threeparttable}
\caption{Optical, near- and mid-IR photometry of the MATLAS NUDGes.}
\begin{tabular}{lccccccccccccccccc} \hline
\multirow{2}{*}{ID} & $g$ & $r$ & $g-r$ & $z$ & $g-z$ & $W1$ & $W2$ & $W3$ & $W4$ \\ 
& [mag] & [mag] & [mag] & [mag] &  [mag] & [mag] & [mag] & [mag] & [mag]  \\ \hline
MATLAS-49 & $19.39 \pm 0.11$ & $18.69 \pm 0.11$ & $0.70$ &  $18.32 \pm 0.11$ & $1.07$ & $18.96 \pm 0.13$ & $20.37 \pm 0.66$ & $>17.41$ & $>14.86$ \\
MATLAS-138 & $18.55 \pm 0.11$ & $17.88 \pm 0.11$ & $0.67$ &  $17.48 \pm 0.11$ & $1.07$ & -- & -- & -- & -- \\
MATLAS-203 & $20.55 \pm 0.13$ & $20.08 \pm 0.19$ & $0.47$ &  -- & -- & $20.22 \pm 0.10$ & $>18.98$ & $>17.05$ & $>14.21$ \\
MATLAS-207 & $20.53 \pm 0.13$ & $19.99 \pm 0.10$ & $0.54$ &  -- & -- & $20.27 \pm 0.14$ & -- & -- & -- \\
MATLAS-290 & $18.71 \pm 0.11$ & $18.13 \pm 0.11$ & $0.59$ &  $17.80 \pm 0.12$ & $0.91$ & -- & -- & -- & -- \\
MATLAS-347 & $17.65 \pm 0.11$ & $17.06 \pm 0.11$ & $0.59$ &  $17.11 \pm 0.10$ & $0.54$ & $18.49 \pm 0.16$ & $18.81 \pm 0.14$ & $17.76 \pm 0.82$ & -- \\
MATLAS-401 & $17.78 \pm 0.10$ & $17.24 \pm 0.11$ & $0.54$ &  $16.88 \pm 0.11$ & $0.90$ & -- & -- & -- & -- \\
MATLAS-524 & $18.98 \pm 0.12$ & $18.31 \pm 0.12$ & $0.67$ &  $18.02 \pm 0.13$ & $0.96$ & $17.92 \pm 0.16$ & $18.00 \pm 0.13$ & $>17.29$ & $>15.15$ \\
MATLAS-627 & $20.13 \pm 0.12$ & $19.45 \pm 0.12$ & $0.68$ &  $19.37 \pm 0.16$ & $0.75$ & $20.37 \pm 0.18$ & -- & $20.78 \pm 1.29$ & -- \\
MATLAS-682 & $21.83 \pm 0.14$ & $21.49 \pm 0.14$ & $0.34$ &  $20.85 \pm 0.18$ & $0.98$ & -- & -- & -- & -- \\
MATLAS-787 & $19.75 \pm 0.13$ & $19.22 \pm 0.13$ & $0.53$ &  $18.79 \pm 0.14$ & $0.96$ & $20.09 \pm 0.19$ & -- & -- & -- \\
MATLAS-791 & $20.03 \pm 0.13$ & $19.39 \pm 0.12$ & $0.65$ &  $19.03 \pm 0.14$ & $1.01$ & $20.03 \pm 0.12$ & -- & -- & -- \\
MATLAS-976 & $19.73 \pm 0.12$ & $19.16 \pm 0.12$ & $0.57$ &  $18.82 \pm 0.12$ & $0.92$ & -- & -- & -- & -- \\
MATLAS-987 & $19.52 \pm 0.13$ & $18.77 \pm 0.12$ & $0.75$ &  $18.34 \pm 0.12$ & $1.18$ & $18.94 \pm 0.16$ & $19.90 \pm 0.54$ & $>17.18$ & $>15.12$ \\
MATLAS-1154 & $18.84 \pm 0.11$ & $18.24 \pm 0.12$ & $0.61$ &  $18.26 \pm 0.10$ & $0.59$ & $18.29 \pm 0.11$ & $18.96 \pm 0.16$ & $>17.79$ & $>15.51$ \\
MATLAS-1321 & $18.23 \pm 0.11$ & $17.60 \pm 0.11$ & $0.63$ &  $17.24 \pm 0.11$ & $0.99$ & -- & -- & -- & -- \\
MATLAS-1332 & $17.52 \pm 0.11$ & $16.88 \pm 0.11$ & $0.63$ &  $16.52 \pm 0.11$ & $1.00$ & $17.03 \pm 0.15$ & $18.11 \pm 0.12$ & $>17.38$ & $>14.92$ \\
MATLAS-1400 & $17.15 \pm 0.11$ & $16.49 \pm 0.11$ & $0.66$ &  $16.14 \pm 0.11$ & $1.01$ & $17.22 \pm 0.14$ & $17.80 \pm 0.09$ & $17.20 \pm 0.77$ & $>14.47$ \\
MATLAS-1408 & $18.64 \pm 0.11$ & $18.08 \pm 0.11$ & $0.56$ &  $17.80 \pm 0.11$ & $0.84$ & -- & -- & -- & -- \\
MATLAS-1412 & $18.13 \pm 0.11$ & $17.56 \pm 0.11$ & $0.57$ &  $17.39 \pm 0.12$ & $0.75$ & $18.79 \pm 0.17$ & $19.67 \pm 0.40$ & $>16.79$ & $>14.57$ \\
MATLAS-1437 & $17.94 \pm 0.11$ & $17.39 \pm 0.11$ & $0.55$ &  $16.78 \pm 0.12$ & $1.16$ & $17.79 \pm 0.12$ & $18.62 \pm 0.19$ & $>17.61$ & $15.21 \pm 1.73$ \\
MATLAS-1470 & $16.93 \pm 0.10$ & $16.40 \pm 0.10$ & $0.53$ &  $16.06 \pm 0.11$ & $0.87$ & -- & -- & -- & -- \\
MATLAS-1485 & $17.42 \pm 0.10$ & $16.87 \pm 0.10$ & $0.54$ &  $16.44 \pm 0.11$ & $0.98$ & $17.51 \pm 0.18$ & $18.11 \pm 0.12$ & -- & -- \\
MATLAS-1530 & $21.35 \pm 0.10$ & $20.43 \pm 0.10$ & $0.91$ &  $20.25 \pm 0.19$ & $1.09$ & $20.04 \pm 0.10$ & $20.96 \pm 0.34$ & -- & -- \\
MATLAS-1539 & $18.91 \pm 0.16$ & $18.34 \pm 0.16$ & $0.57$ &  $18.04 \pm 0.24$ & $0.86$ & $18.89 \pm 0.19$ & $19.52 \pm 0.21$ & $>17.68$ & $>15.02$ \\
MATLAS-1545 & $20.44 \pm 0.16$ & $19.78 \pm 0.16$ & $0.66$ &  $19.42 \pm 0.12$ & $1.02$ & $20.02 \pm 0.19$ & -- & -- & -- \\
MATLAS-1577 & $20.48 \pm 0.12$ & $19.69 \pm 0.12$ & $0.79$ &  $19.09 \pm 0.13$ & $1.39$ & $19.84 \pm 0.14$ & -- & -- & -- \\
MATLAS-1618 & $20.61 \pm 0.16$ & $19.89 \pm 0.15$ & $0.72$ &  $19.77 \pm 0.18$ & $0.84$ & $20.86 \pm 0.12$ & -- & -- & -- \\
MATLAS-1662 & $19.96 \pm 0.11$ & $19.25 \pm 0.11$ & $0.71$ &  $19.01 \pm 0.12$ & $0.96$ & $20.11 \pm 0.10$ & $20.59 \pm 0.95$ & -- & $16.61 \pm 1.51$ \\
MATLAS-1667 & $19.02 \pm 0.12$ & $18.38 \pm 0.11$ & $0.64$ &  $17.97 \pm 0.14$ & $1.05$ & $18.87 \pm 0.12$ & $20.03 \pm 0.23$ & $19.74 \pm 0.74$ & $>15.97$ \\
MATLAS-1740 & $20.48 \pm 0.13$ & $19.86 \pm 0.12$ & $0.62$ &  $19.94 \pm 0.16$ & $0.54$ & $20.81 \pm 0.11$ & $21.57 \pm 0.35$ & -- & -- \\
MATLAS-1801 & $20.56 \pm 0.12$ & $19.84 \pm 0.12$ & $0.71$ &  $19.39 \pm 0.14$ & $1.17$ & $20.82 \pm 0.16$ & -- & -- & -- \\
MATLAS-1888 & $19.33 \pm 0.12$ & $18.74 \pm 0.13$ & $0.59$ &  $18.61 \pm 0.10$ & $0.72$ & $20.03 \pm 0.19$ & $20.53 \pm 0.49$ & $18.07 \pm 0.72$ & -- \\
MATLAS-1938 & $16.55 \pm 0.10$ & $15.88 \pm 0.10$ & $0.67$ &  $15.46 \pm 0.10$ & $1.10$ & $16.55 \pm 0.14$ & $17.33 \pm 0.06$ & $>17.83$ & $>15.88$ \\
MATLAS-2069 & $20.17 \pm 0.13$ & $19.61 \pm 0.13$ & $0.56$ &  $19.45 \pm 0.17$ & $0.72$ & -- & -- & -- & -- \\
MATLAS-2176 & $16.98 \pm 0.14$ & $16.33 \pm 0.14$ & $0.65$ &  $15.89 \pm 0.14$ & $1.09$ & $17.48 \pm 0.15$ & $18.43 \pm 0.12$ & -- & -- \\
\hline
\end{tabular}
\begin{tablenotes}
      \small
      \item \textbf{Note.} Columns are: (1) Galaxy ID; (2) \texttt{GALFITM} DECaLS $g$-band magnitude; (3) \texttt{GALFITM} DECaLS $r$-band magnitude; (4) $g-r$ colour; (5) \texttt{GALFITM} DECaLS $i$-band magnitude; (6) $g-i$ colour; (7) \texttt{GALFITM} DECaLS $z$-band magnitude; (8) $g-z$ colour; (9) \textit{WISE} $3.4 \mu$-band magnitude; (10) \textit{WISE} $4.6 \mu$-band magnitude; (11) \textit{WISE} $12 \mu$-band magnitude; (12) \textit{WISE} $22 \mu$-band magnitude. '--' stands for unavailable data. `>' denotes upper limit magnitudes.
\end{tablenotes}
\label{tab:photometry}
\end{threeparttable}}
\end{table*}

\begin{table*}
\scalebox{0.87}{
\begin{threeparttable}
\caption{\texttt{GALFITM} structural parameters and physical properties of the MATLAS NUDGes and the refitted galaxies from \protect\citetalias{Buzzo_22b}.}
\begin{tabular}{lccccccccccccccccc} \hline
\multirow{2}{*}{ID} & $R_{\rm e}$ & $n$ & $b/a$ & PA & $R_{\rm e}$ & $\langle \mu_{g,\textrm{e}} \rangle$ & $\mu_{g,0}$ & Candidate Host & Host Distance & \texttt{KMeans} Class \\ 
& [arcsec] & & & [degrees] & [kpc] & [mag/arcsec$^2$] & [mag/arcsec$^2$] & & [Mpc] & \\ \hline
MATLAS-49 & $7.66 \pm 0.33$ & $0.82 \pm 0.01$ & $0.44 \pm 0.00$ &$ 79.11 \pm 0.38$ & 1.31 & 24.92 & 24.08 & NGC0502 & 35.9 & A \\
MATLAS-138 & $9.52 \pm 0.33$ & $0.71 \pm 0.01$ & $0.76 \pm 0.00$ & $-46.49 \pm 1.07$ & 1.70 & 25.14 & 24.49 & NGC0680 & 37.5 & B \\
MATLAS-203 & $6.56 \pm 0.54$ & $0.54 \pm 0.03$ & $0.36 \pm 0.01$ & $-15.84 \pm 0.68$ & 1.11 & 25.51 & 25.11 & NGC1121 & 35.3 & A \\
MATLAS-207 & $5.56 \pm 0.50$ & $0.67 \pm 0.03$ & $0.56 \pm 0.01$ & $-14.15 \pm 1.30$ & 0.94 & 25.62 & 25.03 & NGC1121 & 35.3 & A \\
MATLAS-290 & $8.62 \pm 0.47$ & $1.26 \pm 0.02$ & $0.69 \pm 0.00$ & $-24.89 \pm 0.84$ & 1.58 & 24.98 & 23.40 & NGC1289 & 38.4 & A \\
MATLAS-347 & $22.27 \pm 1.13$ & $1.26 \pm 0.01$ & $0.51 \pm 0.00$ & $-32.03 \pm 0.29$ & 1.32 & 25.66 & 24.09 & NGC2549 & 12.3 & A \\
MATLAS-401 & $10.93 \pm 0.45$ & $0.60 \pm 0.01$ & $0.55 \pm 0.08$ &$ 86.64 \pm 0.27$ & 1.83 & 24.32 & 23.83 & NGC2594 & 35.1 & B \\
MATLAS-524 & $12.46 \pm 0.93$ & $0.82 \pm 0.02$ & $0.75 \pm 0.01$ & $-17.05 \pm 1.80$ & 1.61 & 26.14 & 25.32 & NGC2859 & 27.0 & A \\
MATLAS-627 & $7.80 \pm 0.85$ & $1.14 \pm 0.03$ & $0.51 \pm 0.01$ & $-79.67 \pm 3.98 $& 1.70 & 25.86 & 24.50 & UGC05408 & 45.8 & A \\
MATLAS-682 & $6.29 \pm 1.21$ & $0.89 \pm 0.05$ & $0.50 \pm 0.01$ & $-62.42 \pm 1.98$ & 1.22 & 27.07 & 26.13 & NGC3230 & 40.8 & A \\
MATLAS-787 & $9.70 \pm 0.81$ & $0.40 \pm 0.02$ & $0.79 \pm 0.02$ &$ -8.32 \pm 3.57$ & 1.14 & 26.42 & 26.19 & NGC3414 & 24.5 & A \\
MATLAS-791 & $9.00 \pm 0.82$ & $0.52 \pm 0.02$ & $0.68 \pm 0.01$ &$ 10.49 \pm 2.21$ & 1.06 & 26.38 & 25.99 & NGC3414 & 24.5 & A \\
MATLAS-976 & $9.62 \pm 0.75$ & $0.85 \pm 0.02$ & $0.49 \pm 0.01$ &$ 85.51 \pm 0.77$ & 1.21 & 25.87 & 25.00 & NGC3640 & 26.3 & A \\
MATLAS-987 & $8.45 \pm 0.46$ & $0.88 \pm 0.02$ & $0.83 \pm 0.01$ &$ 57.85 \pm 2.19$ & 1.32 & 25.95 & 25.03 & NGC3658 & 32.7 & A \\
MATLAS-1154 & $12.45 \pm 0.55$ & $0.69 \pm 0.01$ & $0.90 \pm 0.01$ &$ 26.75 \pm 3.14$ & 1.47 & 26.19 & 25.57 & NGC4036 & 24.6 & A \\
MATLAS-1321 & $11.07 \pm 0.29$ & $0.88 \pm 0.01$ & $0.66 \pm 0.00$ & $-63.15 \pm 0.44$ & 1.96 & 25.00 & 24.08 & NGC4259 & 37.2 & B \\
MATLAS-1332 & $9.66 \pm 0.24$ & $1.23 \pm 0.01$ & $0.92 \pm 0.00$ &$ 51.12 \pm 1.58$ & 1.71 & 24.35 & 22.83 & NGC4259 & 37.2 & B \\
MATLAS-1400 & $15.61 \pm 0.41$ & $1.43 \pm 0.01$ & $0.55 \pm 0.00$ &$ 59.29 \pm 0.18$ & 1.31 & 24.46 & 22.58 & NGC4623 & 17.4 & A \\
MATLAS-1408 & $8.67 \pm 0.32$ & $0.86 \pm 0.01$ & $0.69 \pm 0.00$ & $-64.88 \pm 71$ & 1.68 & 24.92 & 24.02 & -- & 14.3 & A \\
MATLAS-1412 & $15.82 \pm 0.48$ & $0.95 \pm 0.01$ & $0.88 \pm 0.00$ &$ 61.02 \pm 1.21$ & 1.26 & 25.98 & 24.94 & NGC4643 & 16.5 & A \\
MATLAS-1437 & $17.59 \pm 0.39$ & $0.80 \pm 0.01$ & $0.92 \pm 0.00$ & $-43.01 \pm 1.48$ & 1.40 & 26.07 & 25.28 & NGC4643 & 16.5 & A \\
MATLAS-1470 & $14.24 \pm 0.12$ & $0.98 \pm 0.00$ & $0.97 \pm 0.00$ &$ -1.74 \pm 1.45$ & 1.13 & 24.66 & 23.56 & NGC4643 & 16.5 & A \\
MATLAS-1485 & $15.79 \pm 0.18$ & $1.04 \pm 0.00$ & $0.70 \pm 0.00$ & $-10.70 \pm 0.17$ & 1.25 & 25.01 & 23.82 & NGC4624 & 16.5 & A \\
MATLAS-1530 & $8.60 \pm 3.88$ & $1.14 \pm 0.12$ & $0.69 \pm 0.03$ &$ 59.18 \pm 6.11$ & 1.62 & 27.61 & 26.25 & NGC5198 & 39.6 & A \\
MATLAS-1539 & $8.12 \pm 4.96$ & $2.37 \pm 0.13$ & $0.93 \pm 0.01$ &$ 76.25 \pm 1.16$ & 1.53 & 25.38 & 21.70 & NGC5198 & 39.6 & A \\
MATLAS-1545 & $9.79 \pm 2.84$ & $1.33 \pm 0.09$ & $0.78 \pm 0.03$ &$ 35.17 \pm 6.02$ & 1.85 & 27.12 & 25.42 & NGC5198 & 39.6 & A \\
MATLAS-1577 & $7.54 \pm 0.48$ & $0.52 \pm 0.02$ & $0.69 \pm 0.01$ &$ 70.95 \pm 1.64$ & 1.14 & 26.46 & 26.08 & NGC5308 & 31.5 & A \\
MATLAS-1618 & $7.28 \pm 1.59$ & $1.21 \pm 0.06$ & $1.00 \pm 0.03$ & $0.38 \pm 1.30$ & 1.06 & 26.92 & 25.44 & NGC5322 & 30.3 & A \\
MATLAS-1662 & $7.18 \pm 0.38$ & $0.60 \pm 0.02$ & $0.40 \pm 0.00$ & $-63.67 \pm 0.51$ & 1.27 & 25.24 & 24.75 & NGC5355 & 37.1 & A \\
MATLAS-1667 & $7.83 \pm 0.47$ & $0.78 \pm 0.02$ & $0.90 \pm 0.01$ &$ 22.16 \pm 4.57$ & 1.12 & 25.37 & 24.61 & NGC5379 & 30.0 & A \\
MATLAS-1740 & $10.67 \pm 0.97$ & $0.66 \pm 0.03$ & $0.29 \pm 0.00$ &$ 71.68 \pm 0.53$ & 1.32 & 26.28 & 25.71 & NGC5481 & 25.8 & A \\
MATLAS-1801 & $7.00 \pm 0.65$ & $0.67 \pm 0.02$ & $0.63 \pm 0.01$ & $-45.24 \pm 1.73$ & 1.29 & 26.28 & 25.69 & NGC5557 & 38.8 & A \\
MATLAS-1888 & $11.40 \pm 0.89$ & $0.76 \pm 0.02$ & $0.74 \pm 0.01$ & $4.98 \pm 1.96$ & 1.47 & 26.28 & 25.55 & NGC5631 & 27.0 & A \\
MATLAS-1938 & $12.55 \pm 0.09$ & $0.98 \pm 0.00$ & $0.75 \pm 0.00$ &$ 33.21 \pm 0.14$ & 1.88 & 23.72 & 22.63 & NGC5813 & 31.3 & B \\
MATLAS-2069 & $10.69 \pm 0.87$ & $0.41 \pm 0.03$ & $0.51 \pm 0.01$ & $-57.88 \pm 1.35$ & 1.29 & 26.58 & 26.33 & NGC5845 & 25.2 & A \\
MATLAS-2176 & $17.25 \pm 2.92$ & $0.68 \pm 0.02$ & $1.00 \pm 0.05$ & $-43.62 \pm 2.19$ & 1.92 & 25.16 & 24.56 & NGC7454 & 23.2 & A \\
\hline \hline
DF02 & $3.75 \pm 0.39$ & $0.62 \pm 0.04$ & $0.76 \pm 0.02$ & $-63.71 \pm 3.57$ & 1.74 & 25.31 & 24.79 & Coma cluster & 100.0 & A \\
DF03 & $5.42 \pm 0.52$ & $0.73 \pm 0.03$ & $0.49 \pm 0.01$ & $-87.44 \pm 1.08$ & 3.58 & 25.62 & 24.94 & Group behind Coma & 145.0 & B \\
DF06 & $8.37 \pm 2.01$ & $0.74 \pm 0.05$ & $0.35 \pm 0.01$ & $-52.30 \pm 1.64$ & 3.89 & 26.63 & 25.93 & Coma cluster & 100.0 & A \\
DF07 & $6.68 \pm 0.35$ & $0.69 \pm 0.01$ & $0.78 \pm 0.01$ & $-48.24 \pm 1.79$ & 3.10 & 25.56 & 24.94 & Coma cluster & 100.0 & B \\
DF08 & $4.74 \pm 0.48$ & $0.72 \pm 0.02$ & $0.98 \pm 0.02$ & $-88.97 \pm 1.57$ & 2.20 & 26.25 & 25.58 & Coma cluster & 100.0 & B \\
DF17 & $3.96 \pm 3.14$ & $0.55 \pm 0.42$ & $0.98 \pm 0.02$ & $-64.24 \pm 9.50$ & 1.84 & 25.25 & 24.83 & Coma cluster & 100.0 & B \\
DF23 & $4.80 \pm 0.60$ & $0.92 \pm 0.04$ & $0.88 \pm 0.02$ & $-83.06 \pm 7.49$ & 2.23 & 25.99 & 25.00 & Coma cluster & 100.0& B \\
DF25 & $10.49 \pm 0.51$ & $0.10 \pm 0.01$ & $0.40 \pm 0.01$ & $38.01 \pm 0.64$ & 4.87 & 27.31 & 27.30 & Coma cluster & 100.0 & A \\
DF26 & $6.27 \pm 0.37$ & $0.79 \pm 0.02$ & $0.64 \pm 0.01$ & $-4.36 \pm 1.06$ & 2.91 & 25.69 & 24.91 & Coma cluster & 100.0 & B \\
DF40 & $10.37 \pm 1.78$ & $1.48 \pm 0.06$ & $0.43 \pm 0.01$ & $-51.34 \pm 0.88$ & 4.81 & 26.95 & 24.98 & Coma cluster & 100.0 & A \\
DF44 & $8.19 \pm 0.49$ & $0.77 \pm 0.02$ & $0.67 \pm 0.01$ & $-28.21 \pm 1.19$ & 3.80 & 26.18 & 25.43 & Coma cluster & 100.0 & B \\
DF46 & $4.42 \pm 0.89$ & $0.68 \pm 0.05$ & $0.62 \pm 0.02$ & $-42.74 \pm 3.69$ & 2.05 & 26.12 & 25.51 & Coma cluster & 100.0 & A \\
DFX1 & $7.27 \pm 0.45$ & $0.82 \pm 0.02$ & $0.55 \pm 0.01$ & $9.36 \pm 0.78$ & 3.37 & 25.81 & 24.98 & Coma cluster & 100.0 & B \\
DFX2 & $5.03 \pm 1.10$ & $1.79 \pm 0.10$ & $0.90 \pm 0.90$ & $-5.54 \pm 9.95$ & 2.33 & 25.91 & 23.35 & Coma cluster & 100.0 & A \\
DGSATI & $6.59 \pm 0.89$ & $0.92 \pm 0.56$ & $0.80 \pm 0.10$ & $-33.18 \pm 2.87$ & 2.17 & 24.56 & 23.43 & Isolated & 70.0 & B \\
LSBG-044 & $5.92 \pm 0.31$ & $0.45 \pm 0.02$ & $0.62 \pm 0.01$ & $59.88 \pm 1.12$ & 4.87 & 25.98 & 25.69 & Isolated & -- & -- \\
LSBG-378 & $6.59 \pm 0.46$ & $1.01 \pm 0.01$ & $0.80 \pm 0.10$ & $-17.48 \pm 3.34$ & 4.03 & 25.01 & 23.88 & Isolated & -- & -- \\
LSBG-490 & $4.83 \pm 0.18$ & $1.03 \pm 0.01$ & $0.85 \pm 0.00$ & $53.36 \pm 1.57$ & 3.33 & 25.80 & 24.62 & Isolated & -- & -- \\
N1052-DF2 & $20.97 \pm 0.21$ & $0.55 \pm 0.00$ & $0.89 \pm 0.00$ & $-43.13 \pm 0.54$ & 2.23 & 25.56 & 25.14 & NGC1052 & 22.1 & B \\
N1052-DF4 & $16.06 \pm 0.25$ & $0.79 \pm 0.00$ & $0.87 \pm 0.00$ & $-82.50 \pm 0.68$ & 1.54 & 25.65 & 24.87 & NGC1052 & 20.0 & B \\
PUDG-R16 & $7.91 \pm 1.16$ & $1.35 \pm 0.05$ & $0.90 \pm 0.04$ & $52.55 \pm 8.48$ & 2.78 & 26.49 & 24.76 & Perseus cluster & 75.0 & B \\
PUDG-R24 & $13.66 \pm 0.23$ & $1.00 \pm 0.23$ & $0.83 \pm 0.10$ & $-29.34 \pm 0.69$ & 4.81 & 27.06 & 25.94 & Perseus cluster & 75.0 & A \\
VCC1052 & $66.36 \pm 3.58$ & $0.71 \pm 0.02$ & $0.81 \pm 0.00$ & $16.11 \pm 0.51$ & 5.27 & 28.09 & 27.43 & Virgo cluster & 16.5 & B \\
VCC1287 & $46.70 \pm 8.17$ & $0.72 \pm 0.02$ & $0.86 \pm 0.00$ & $28.15 \pm 1.30$ & 3.71 & 27.35 & 26.67 & Virgo cluster & 16.5 & B \\
VCC1884 & $38.46 \pm 2.92$ & $0.57 \pm 0.01$ & $0.78 \pm 0.00$ & $69.94 \pm 0.70$ & 3.05 & 27.60 & 27.15 & Virgo cluster & 16.5 & A \\
Y358 & $5.54 \pm 1.43$ & $1.21 \pm 0.16$ & $0.71 \pm 0.20$ & $41.05 \pm 6.45$ & 2.57 & 26.12 & 24.65 & Coma cluster & 100.0 & B \\
Y436 & $3.23 \pm 0.42$ & $0.65 \pm 0.06$ & $0.79 \pm 0.02$ & $52.21 \pm 5.04$ & 1.50 & 25.98 & 25.42 & Coma cluster & 100.0 & B \\
Y534 & $6.99 \pm 1.62$ & $1.54 \pm 0.08$ & $0.67 \pm 0.01$ & $-39.97 \pm 2.76$ & 3.24 & 27.38 & 25.30 & Coma cluster & 100.0 & B \\ \hline
\end{tabular}
\begin{tablenotes}
      \small
      \item \textbf{Note.} Columns are: (1) Galaxy ID; (2) \texttt{GALFITM} effective radius in arcsec; (3) \texttt{GALFITM} S\'ersic index; (4) \texttt{GALFITM} axis ratio; (5) \texttt{GALFITM} position angle; (6) Effective radius in kpc; (7) Mean surface brightness; (8) Central surface brightness; (9) Candidate hosts of NUDGes and UDGs; (10) Distance to host (and assumed distance to NUDG and UDG); (11) \texttt{KMeans} clustering algorithm class. 
\end{tablenotes}
\label{tab:morphology}
\end{threeparttable}}
\end{table*}

\begin{table*}
\scalebox{0.85}{
\begin{threeparttable}
\caption{\texttt{Prospector} stellar population properties of the MATLAS NUDGes and the refitted UDGs from \protect\citetalias{Buzzo_22b}.}
\begin{tabular}{lcccccccccccccccccc} \hline
\multirow{2}{*}{ID} & \multirow{2}{*}{log($M_{\star}/M_{\odot}$)} & [M/H] & $\tau$ & $t_M$ & $A_v$ & $M_{\star}/L_{V}$ \\ 
  &  & [dex] & [Gyr] & [Gyr] & [mag] & [$M_{\odot}/L_{\odot,V}$]  \\ \hline
MATLAS-49 & $7.87^{+0.05}_{-0.06}$ & $-0.88^{+0.33}_{-0.47}$ & $0.50^{+0.39}_{-0.28}$ & $11.00^{+1.95}_{-2.79}$ & $0.26^{+0.13}_{-0.12}$ & $2.84$ \\
MATLAS-138 & $8.22^{+0.05}_{-0.07}$ & $-0.92^{+0.37}_{-0.43}$ & $0.55^{+0.46}_{-0.32}$ & $10.93^{+1.97}_{-3.21}$ & $0.27^{+0.12}_{-0.14}$ & $2.72$ \\
MATLAS-203 & $7.02^{+0.30}_{-0.27}$ & $-0.65^{+0.74}_{-0.57}$ & $3.53^{+4.12}_{-2.66}$ & $4.99^{+6.02}_{-2.83}$ & $0.48^{+0.35}_{-0.31}$ & $1.20$ \\
MATLAS-207 & $7.14^{+0.26}_{-0.24}$ & $-0.67^{+0.64}_{-0.57}$ & $3.44^{+3.96}_{-2.53}$ & $5.25^{+5.85}_{-2.87}$ & $0.43^{+0.29}_{-0.26}$ & $1.58$ \\
MATLAS-290 & $8.10^{+0.05}_{-0.07}$ & $-1.11^{+0.38}_{-0.28}$ & $0.66^{+0.54}_{-0.39}$ & $10.40^{+2.41}_{-3.58}$ & $0.15^{+0.09}_{-0.10}$ & $2.25$ \\
MATLAS-347 & $7.27^{+0.05}_{-0.03}$ & $-1.48^{+0.03}_{-0.01}$ & $0.43^{+0.37}_{-0.23}$ & $7.62^{+2.25}_{-1.54}$ & $0.01^{+0.01}_{-0.00}$ & $1.24$ \\
MATLAS-401 & $8.25^{+0.05}_{-0.07}$ & $-1.02^{+0.34}_{-0.33}$ & $1.38^{+0.64}_{-0.84}$ & $8.24^{+3.75}_{-3.91}$ & $0.19^{+0.10}_{-0.11}$ & $1.62$ \\
MATLAS-524 & $7.75^{+0.08}_{-0.09}$ & $0.21^{+0.13}_{-0.22}$ & $2.70^{+1.34}_{-1.06}$ & $6.17^{+3.68}_{-2.92}$ & $0.54^{+0.13}_{-0.13}$ & $2.64$ \\
MATLAS-627 & $7.67^{+0.05}_{-0.06}$ & $-1.20^{+0.35}_{-0.19}$ & $0.54^{+0.48}_{-0.32}$ & $11.28^{+1.81}_{-3.41}$ & $0.11^{+0.08}_{-0.07}$ & $2.17$ \\
MATLAS-682 & $6.94^{+0.19}_{-0.14}$ & $-0.81^{+0.80}_{-0.56}$ & $4.65^{+3.55}_{-3.12}$ & $4.31^{+5.18}_{-2.18}$ & $0.76^{+0.44}_{-0.43}$ & $2.47$ \\
MATLAS-787 & $7.15^{+0.07}_{-0.08}$ & $-1.11^{+0.31}_{-0.27}$ & $0.82^{+0.70}_{-0.51}$ & $9.40^{+3.19}_{-4.03}$ & $0.15^{+0.11}_{-0.10}$ & $1.64$ \\
MATLAS-791 & $7.23^{+0.06}_{-0.08}$ & $-0.98^{+0.39}_{-0.37}$ & $0.69^{+0.63}_{-0.42}$ & $10.03^{+2.66}_{-4.17}$ & $0.25^{+0.13}_{-0.14}$ & $2.54$ \\
MATLAS-976 & $7.36^{+0.08}_{-0.09}$ & $-1.06^{+0.38}_{-0.30}$ & $1.07^{+1.00}_{-0.70}$ & $9.04^{+3.48}_{-4.41}$ & $0.20^{+0.17}_{-0.12}$ & $2.24$ \\
MATLAS-987 & $7.67^{+0.14}_{-0.20}$ & $-0.65^{+0.61}_{-0.58}$ & $1.12^{+1.61}_{-0.73}$ & $8.82^{+3.57}_{-4.99}$ & $0.42^{+0.24}_{-0.26}$ & $2.47$ \\
MATLAS-1154 & $7.40^{+0.07}_{-0.10}$ & $-0.46^{+0.33}_{-0.53}$ & $1.63^{+0.74}_{-0.95}$ & $7.29^{+4.32}_{-4.11}$ & $0.32^{+0.15}_{-0.17}$ & $1.24$ \\
MATLAS-1321 & $8.31^{+0.06}_{-0.06}$ & $-0.73^{+0.19}_{-0.58}$ & $0.54^{+0.39}_{-0.30}$ & $10.64^{+2.23}_{-2.48}$ & $0.12^{+0.16}_{-0.08}$ & $2.53$ \\
MATLAS-1332 & $8.25^{+0.10}_{-0.14}$ & $-1.30^{+0.20}_{-0.20}$ & $0.85^{+0.89}_{-0.54}$ & $9.25^{+3.23}_{-4.62}$ & $0.15^{+0.15}_{-0.11}$ & $1.13$ \\
MATLAS-1400 & $7.70^{+0.15}_{-0.21}$ & $-0.43^{+0.63}_{-0.65}$ & $1.45^{+2.18}_{-0.95}$ & $7.95^{+4.12}_{-4.52}$ & $0.14^{+0.18}_{-0.10}$ & $1.05$ \\
MATLAS-1408 & $8.14^{+0.03}_{-0.04}$ & $-1.26^{+0.16}_{-0.16}$ & $0.50^{+0.42}_{-0.29}$ & $11.71^{+1.41}_{-2.82}$ & $0.08^{+0.06}_{-0.05}$ & $2.05$ \\
MATLAS-1412 & $7.53^{+0.01}_{-0.02}$ & $-1.25^{+0.05}_{-0.06}$ & $0.36^{+0.34}_{-0.19}$ & $12.37^{+0.92}_{-1.28}$ & $0.03^{+0.03}_{-0.02}$ & $1.93$ \\
MATLAS-1437 & $7.59^{+0.02}_{-0.04}$ & $0.18^{+0.06}_{-0.06}$ & $2.28^{+0.35}_{-0.56}$ & $7.90^{+2.19}_{-2.41}$ & $0.02^{+0.03}_{-0.01}$ & $1.88$ \\
MATLAS-1470 & $7.87^{+0.04}_{-0.05}$ & $-1.00^{+0.25}_{-0.27}$ & $1.08^{+0.47}_{-0.61}$ & $8.58^{+3.08}_{-3.50}$ & $0.11^{+0.09}_{-0.07}$ & $1.40$ \\
MATLAS-1485 & $7.65^{+0.03}_{-0.10}$ & $-0.19^{+0.05}_{-0.08}$ & $1.86^{+0.40}_{-0.98}$ & $7.96^{+3.62}_{-4.37}$ & $0.02^{+0.05}_{-0.01}$ & $1.34$ \\
MATLAS-1530 & $7.23^{+0.18}_{-0.24}$ & $-0.64^{+0.62}_{-0.58}$ & $1.69^{+3.47}_{-1.19}$ & $7.66^{+4.39}_{-4.59}$ & $0.96^{+0.18}_{-0.25}$ & $3.24$ \\
MATLAS-1539 & $7.92^{+0.09}_{-0.12}$ & $-1.12^{+0.39}_{-0.32}$ & $1.00^{+1.03}_{-0.63}$ & $9.01^{+3.40}_{-4.70}$ & $0.16^{+0.11}_{-0.10}$ & $1.70$ \\
MATLAS-1545 & $7.34^{+0.11}_{-0.14}$ & $-0.86^{+0.53}_{-0.44}$ & $1.23^{+1.63}_{-0.81}$ & $8.51^{+3.71}_{-4.66}$ & $0.38^{+0.20}_{-0.22}$ & $1.83$ \\
MATLAS-1577 & $7.42^{+0.07}_{-0.08}$ & $-0.85^{+0.48}_{-0.44}$ & $1.22^{+1.24}_{-0.78}$ & $8.80^{+3.47}_{-4.31}$ & $0.75^{+0.22}_{-0.24}$ & $3.59$ \\
MATLAS-1618 & $7.15^{+0.08}_{-0.10}$ & $-1.12^{+0.39}_{-0.26}$ & $0.71^{+0.65}_{-0.43}$ & $9.97^{+2.75}_{-4.31}$ & $0.16^{+0.13}_{-0.10}$ & $2.35$ \\
MATLAS-1662 & $7.63^{+0.05}_{-0.07}$ & $-1.06^{+0.49}_{-0.26}$ & $0.48^{+0.41}_{-0.27}$ & $11.36^{+1.70}_{-3.09}$ & $0.20^{+0.09}_{-0.14}$ & $2.58$ \\
MATLAS-1667 & $7.74^{+0.05}_{-0.06}$ & $-0.70^{+0.18}_{-0.27}$ & $0.55^{+0.44}_{-0.32}$ & $10.97^{+2.04}_{-3.01}$ & $0.12^{+0.09}_{-0.07}$ & $2.14$ \\
MATLAS-1740 & $6.72^{+0.12}_{-0.16}$ & $-0.79^{+0.62}_{-0.48}$ & $5.73^{+2.93}_{-3.03}$ & $4.29^{+3.57}_{-1.90}$ & $0.37^{+0.18}_{-0.19}$ & $1.89$ \\
MATLAS-1801 & $7.47^{+0.05}_{-0.06}$ & $-1.12^{+0.43}_{-0.24}$ & $0.53^{+0.45}_{-0.30}$ & $11.36^{+1.73}_{-2.96}$ & $0.30^{+0.09}_{-0.14}$ & $2.86$ \\
MATLAS-1888 & $7.49^{+0.06}_{-0.07}$ & $-1.23^{+0.30}_{-0.19}$ & $0.64^{+0.57}_{-0.38}$ & $10.30^{+2.51}_{-3.60}$ & $0.11^{+0.09}_{-0.07}$ & $1.97$ \\
MATLAS-1938 & $8.89^{+0.00}_{-0.00}$ & $-1.24^{+0.01}_{-0.01}$ & $0.13^{+0.05}_{-0.03}$ & $12.59^{+0.20}_{-0.22}$ & $0.34^{+0.00}_{-0.00}$ & $2.85$ \\
MATLAS-2069 & $7.06^{+0.08}_{-0.10}$ & $-1.20^{+0.37}_{-0.21}$ & $1.05^{+1.16}_{-0.67}$ & $8.95^{+3.46}_{-4.54}$ & $0.12^{+0.16}_{-0.08}$ & $1.86$ \\
MATLAS-2176 & $8.12^{+0.06}_{-0.05}$ & $-1.45^{+0.08}_{-0.04}$ & $0.61^{+0.56}_{-0.36}$ & $7.28^{+3.36}_{-2.41}$ & $0.01^{+0.01}_{-0.01}$ & $1.31$ \\
  \hline \hline
DF02 & $8.03^{+0.19}_{-0.18}$ & $-1.01^{+0.20}_{-0.20}$ & $4.46^{+3.66}_{-2.99}$ & $3.95^{+4.79}_{-1.67}$ & $0.67^{+0.39}_{-0.38}$ & 1.85 \\
DF03 & $8.13^{+0.22}_{-0.23}$ & $-1.47^{+0.56}_{-0.37}$ & $2.05^{+3.48}_{-1.48}$ & $6.78^{+5.10}_{-4.14}$ & $0.14^{+0.14}_{-0.10}$ & 1.11 \\
DF06 & $7.62^{+0.25}_{-0.19}$ & $-1.21^{+0.30}_{-0.30}$ & $4.29^{+3.54}_{-3.01}$ & $6.50^{+6.43}_{-2.51}$ & $0.50^{+0.25}_{-0.23}$ & 1.07 \\
DF07 & $8.58^{+0.12}_{-0.17}$ & $-1.25^{+0.31}_{-0.29}$ & $0.96^{+1.03}_{-0.62}$ & $9.14^{+3.46}_{-4.84}$ & $0.56^{+0.26}_{-0.22}$ & 2.51 \\
DF08 & $8.22^{+0.18}_{-0.20}$ & $-1.25^{+0.29}_{-0.29}$ & $2.51^{+3.83}_{-1.76}$ & $8.36^{+5.86}_{-3.92}$ & $0.30^{+0.27}_{-0.20}$ & 3.27 \\
DF17 & $8.17^{+0.17}_{-0.19}$ & $-1.63^{+0.42}_{-0.27}$ & $1.85^{+3.13}_{-1.28}$ & $6.99^{+4.75}_{-4.15}$ & $0.06^{+0.07}_{-0.05}$ & 1.66 \\
DF23 & $8.12^{+0.18}_{-0.21}$ & $-1.54^{+0.28}_{-0.28}$ & $2.14^{+3.60}_{-1.42}$ & $8.21^{+5.23}_{-4.10}$ & $0.25^{+0.14}_{-0.17}$ & 2.27 \\
DF25 & $7.96^{+0.21}_{-0.24}$ & $-1.21^{+0.76}_{-0.61}$ & $4.51^{+3.58}_{-2.94}$ & $4.10^{+5.75}_{-2.65}$ & $0.88^{+0.52}_{-0.48}$ & 2.39 \\
DF26 & $8.34^{+0.14}_{-0.20}$ & $-1.19^{+0.34}_{-0.22}$ & $1.31^{+1.88}_{-0.88}$ & $8.18^{+3.99}_{-4.78}$ & $0.14^{+0.13}_{-0.09}$ & 2.29 \\
DF40 & $7.90^{+0.10}_{-0.20}$ & $-0.70^{+0.66}_{-0.82}$ & $4.90^{+3.23}_{-2.98}$ & $4.64^{+4.45}_{-2.36}$ & $0.57^{+0.22}_{-0.25}$ & 1.45 \\
DF44 & $8.39^{+0.11}_{-0.14}$ & $-1.51^{+0.20}_{-0.19}$ & $1.18^{+1.41}_{-0.73}$ & $9.06^{+4.31}_{-4.42}$ & $0.47^{+0.12}_{-0.12}$ & 2.22 \\
DF46 & $7.62^{+0.27}_{-0.21}$ & $-1.05^{+0.28}_{-0.29}$ & $4.04^{+6.67}_{-3.02}$ & $4.47^{+5.51}_{-1.92}$ & $0.23^{+0.24}_{-0.16}$ & 1.34 \\
DFX1 & $8.31^{+0.11}_{-0.14}$ & $-1.41^{+0.25}_{-0.26}$ & $1.42^{+1.52}_{-0.93}$ & $9.27^{+4.80}_{-4.48}$ & $0.10^{+0.10}_{-0.07}$ & 2.03 \\
DFX2 & $7.98^{+0.10}_{-0.12}$ & $-0.75^{+0.18}_{-0.19}$ & $6.92^{+2.13}_{-2.77}$ & $2.73^{+1.66}_{-0.96}$ & $0.20^{+0.05}_{-0.05}$ & 1.02 \\
DGSATI & $8.61^{+0.05}_{-0.10}$ & $-1.81^{+0.13}_{-0.09}$ & $0.47^{+0.50}_{-0.27}$ & $10.38^{+3.09}_{-4.14}$ & $0.22^{+0.02}_{-0.02}$ & 2.19 \\
LSBG-044 & $7.94^{+0.27}_{-0.31}$ & $-1.04^{+0.53}_{-0.61}$ & $1.79^{+3.42}_{-1.17}$ & $7.63^{+4.37}_{-4.42}$ & $0.24^{+0.18}_{-0.15}$ & -- \\
LSBG-378 & $8.98^{+0.31}_{-0.77}$ & $-0.87^{+0.66}_{-0.76}$ & $1.76^{+3.23}_{-1.25}$ & $6.94^{+4.90}_{-3.79}$ & $0.21^{+0.23}_{-0.15}$ & -- \\
LSBG-490 & $8.18^{+0.24}_{-0.19}$ & $-1.03^{+0.64}_{-0.66}$ & $2.55^{+3.23}_{-1.67}$ & $6.67^{+4.73}_{-3.74}$ & $0.48^{+0.22}_{-0.24}$ & -- \\
N1052-DF2 & $8.18^{+0.09}_{-0.14}$ & $-1.02^{+0.29}_{-0.29}$ & $0.76^{+0.81}_{-0.47}$ & $9.68^{+2.93}_{-4.64}$ & $0.02^{+0.03}_{-0.02}$ & 1.84 \\
N1052-DF4 & $8.05^{+0.06}_{-0.10}$ & $-0.89^{+0.19}_{-0.21}$ & $0.69^{+0.59}_{-0.40}$ & $10.26^{+2.51}_{-4.02}$ & $0.11^{+0.09}_{-0.07}$ & 3.11 \\
PUDG-R16 & $8.09^{+0.15}_{-0.23}$ & $-0.96^{+0.26}_{-0.25}$ & $1.25^{+1.87}_{-0.83}$ & $8.31^{+3.92}_{-4.75}$ & $0.14^{+0.14}_{-0.10}$ & 1.92 \\
PUDG-R24 & $8.23^{+0.14}_{-0.17}$ & $-1.16^{+0.24}_{-0.25}$ & $0.96^{+1.68}_{-0.64}$ & $4.13^{+3.61}_{-2.46}$ & $0.05^{+0.07}_{-0.04}$ & 1.86 \\
VCC1052 & $7.73^{+0.12}_{-0.10}$ & $-1.22^{+0.24}_{-0.25}$ & $0.64^{+0.71}_{-0.39}$ & $11.25^{+2.32}_{-3.09}$ & $0.09^{+0.11}_{-0.09}$ & 3.34 \\
VCC1287 & $7.97^{+0.08}_{-0.12}$ & $-1.43^{+0.38}_{-0.34}$ & $0.76^{+0.72}_{-0.46}$ & $9.69^{+2.91}_{-4.33}$ & $0.01^{+0.02}_{-0.01}$ & 2.18 \\
VCC1884 & $7.46^{+0.11}_{-0.07}$ & $-1.48^{+0.23}_{-0.26}$ & $0.76^{+3.13}_{-2.31}$ & $13.02^{+3.55}_{-2.20}$ & $0.11^{+0.21}_{-0.11}$ & 1.39 \\
Y358 & $8.22^{+0.20}_{-0.18}$ & $-1.47^{+0.30}_{-0.30}$ & $3.05^{+3.21}_{-2.01}$ & $6.99^{+5.79}_{-3.05}$ & $0.74^{+0.15}_{-0.15}$ & 2.99 \\
Y436 & $7.90^{+0.24}_{-0.24}$ & $-1.29^{+0.31}_{-0.30}$ & $2.96^{+4.63}_{-2.18}$ & $7.49^{+6.34}_{-3.44}$ & $0.44^{+0.05}_{-0.10}$ & 3.32 \\
Y534 & $7.94^{+0.16}_{-0.21}$ & $-1.52^{+0.29}_{-0.29}$ & $1.93^{+2.87}_{-1.33}$ & $8.53^{+5.17}_{-4.20}$ & $0.35^{+0.19}_{-0.17}$ & 3.35 \\ \hline
\end{tabular}
\begin{tablenotes}
      \small
      \item \textbf{Note.} Columns are: (1) Galaxy ID; (2) \texttt{Prospector} stellar mass; (3) \texttt{Prospector} metallicity; (4) \texttt{Prospector} Star formation timescale; (5)  \texttt{Prospector} Mass-weighted Age; (6) \texttt{Prospector} Dust attenuation; (7) Mass-to-light ratio. 
\end{tablenotes}
\label{tab:stellarpops}
\end{threeparttable}}
\end{table*}

\section{Correlations between multiple properties of UDGs and NUDGes}
\label{sec:heatmap}
\begin{figure*}
    \centering
    \includegraphics[width=0.8\textwidth]{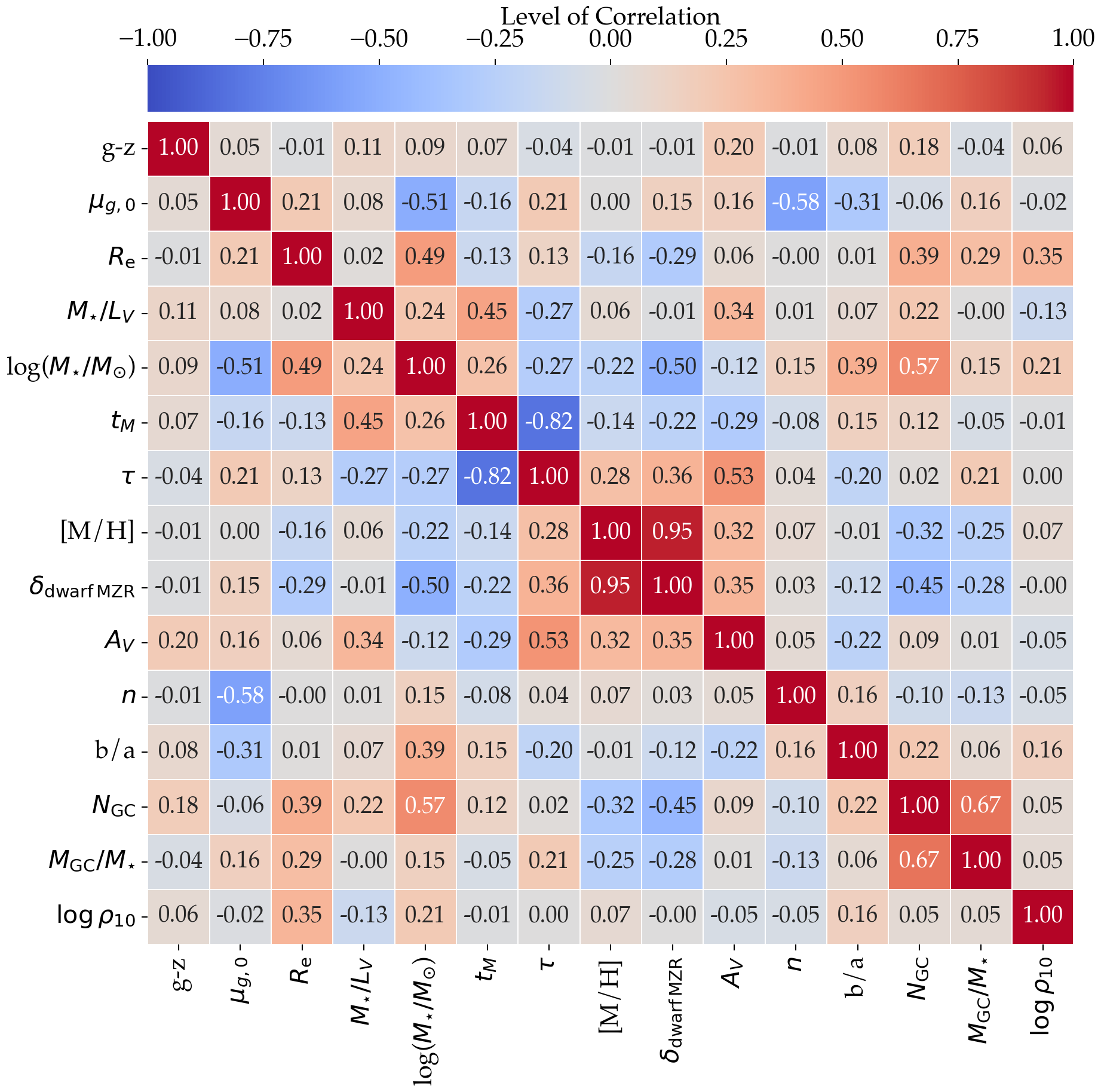}
    \caption{Correlation matrix showing the level of correlation between the different physical properties of the galaxies (UDGs+NUDGes) in our sample. The level of correlation is traced by a continuous colour map, with dark red being the highest level of correlation and dark blue being the highest level of anti-correlation. Properties that are not correlated have white colours. The number on each square provides a quantitative measure of the level of correlation between properties.}
    \label{fig:heatmap}
\end{figure*}

In this Appendix, we examine the correlations between various properties of the galaxies in our sample. 
Figure \ref{fig:heatmap} presents a heatmap that highlights the correlations among different morphological, physical, and stellar population properties of the galaxies (UDGs + NUDGes).
The correlation levels in the heatmap are derived from a correlation matrix, which quantifies the linear relationships between variables by computing pairwise Pearson correlation coefficients. Each coefficient ranges between $-1$ and 1, where values closer to 1 indicate a strong positive linear correlation, values near 0 suggest no linear relationship, and values closer to $-1$ signify a strong negative linear correlation. 
In our particular case, we define strong positive correlations as those with a Correlation Level $\geq$ 0.5, while weaker positive correlations are identified with 0.25 < Correlation Levels < 0.5. Values between $-0.25$ and 0.25 indicate negligible correlations. Weak negative correlations range from $-0.5$ to $-0.25$, and strong negative correlations are identified by Correlation Levels $\leq -0.5$. While some correlations in Fig. \ref{fig:heatmap} are well-known and expected across all galaxy types, others are new and may provide intriguing insights into the formation scenarios for low surface brightness dwarf galaxies, such as UDGs and NUDGes.

A detailed examination of Fig. \ref{fig:heatmap} reveals that the colour index $g-z$ does not exhibit strong correlations with any other analysed parameter. The central surface brightness $\mu_0$ shows strong anti-correlations with both stellar mass and S\'ersic index, suggesting that brighter galaxies tend to be more massive and possess higher S\'ersic indices, as expected. A somewhat weaker anti-correlation is observed between surface brightness and axis ratio, indicating that brighter galaxies tend to be rounder, consistent with the recent findings of \cite{Pfeffer_24}, who noted that brighter dwarf galaxies host more globular clusters (GCs) and are rounder in shape.

The effective radius $R_e$ strongly correlates with stellar mass, a relationship previously discussed in several studies (e.g., \cite{Harris_13}). Additionally, $R_e$ exhibits a slight anti-correlation with $\delta_{\rm dwarf \, MZR}$ and a positive correlation with the number of GCs, implying that the largest galaxies in our sample tend to host more GCs and lie below the classical dwarf mass-metallicity relation (MZR), as previously discussed in Section \ref{sec:classes}.

The mass-to-light ratio ($M_{\star}/L_V$) correlates with both age and dust content, suggesting that galaxies with higher $M_{\star}/L_V$ ratios are generally older and contain more dust. While the age behaviour is expected and discussed in Section \ref{sec:classes}, the dust correlation is less explored in prior works. However, it is important to note that the maximum dust content observed in these galaxies is 0.88 mag, which remains relatively low and insignificant. Additionally, the dust component added by \texttt{PROSPECTOR} might be an artificial addition to improve the model fits.

Stellar mass correlates with nearly every other property, as extensively discussed in the literature. These correlations are why many properties in the clustering algorithm are normalised by mass. The mass-weighted age ($t_M$) shows a strong anti-correlation with the star formation timescale ($\tau$), as expected. Moreover, $\tau$ strongly correlates with metallicity, the distance to the classical dwarf MZR, and dust attenuation, indicating that galaxies with rapid star formation histories (low $\tau$) tend to be more metal-poor and have lower dust content, aligning with theoretical predictions.

Furthermore, metallicity [M/H] shows strong correlations with the distance to the MZR ($\delta_{\rm dwarf \, MZR}$) by definition, as well as with dust content. It is also strongly anti-correlated with the GC number, indicating that the most GC-rich galaxies are the most metal-poor, as discussed in Section \ref{sec:classes}. All remaining correlations have been previously discussed. The key interpretations of these correlations are presented in the main body of the paper, particularly in Sections \ref{sec:classes} and \ref{sec:udg_definition}.

\bsp	
\label{lastpage}
\end{document}